\PassOptionsToPackage{dvipsnames}{xcolor}

\documentclass[10pt, conference]{IEEEtran}

\usepackage{cite}

\ifCLASSINFOpdf
   \usepackage[pdftex]{graphicx}
   \graphicspath{{./figs/}{./pics/}}
   \DeclareGraphicsExtensions{.pdf,.jpeg,.png}
\else

\fi

\newif\ifMiKTeX
\makeatletter
\def\testMiKTeX{\expandafter\testMiKTeX@i\pdftexbanner\@nil}
\def\testMiKTeX@i#1(#2)#3\@nil{\testMiKTeX@ii#2..\@nil}
\def\testMiKTeX@ii#1.#2.#3\@nil{\ifx\relax#2\relax\else\MiKTeXtrue\fi}
\makeatother
\testMiKTeX

\usepackage{tikz}
\usetikzlibrary{shapes.geometric, arrows}

\usepackage[cmex10]{amsmath}
\usepackage[linesnumbered,ruled,vlined]{algorithm2e}
\usepackage{algpseudocode}
\usepackage{array}
\usepackage{amssymb}

\ifCLASSOPTIONcompsoc
  \usepackage[caption=false,font=normalsize,labelfont=sf,textfont=sf]{subfig}
\else
  \usepackage[caption=false,font=footnotesize]{subfig}
\fi

\usepackage{url}

\usepackage{units}
\usepackage{enumitem}

\usepackage[dvipsnames]{xcolor}
\usepackage{hhline}

\definecolor{myColor}{HTML}{000000}

\usepackage{gensymb}
\usepackage{currvita}

\usepackage{fixltx2e}
\usepackage{amsmath}
\usepackage[bottom]{footmisc}
\usepackage{textcomp}
\usepackage{xfrac}

\usepackage{mathtools}

\begin{document}


\title{\huge{FPGA Energy Efficiency by Leveraging Thermal Margin}}

\author{Behnam Khaleghi, Sahand Salamat, Mohsen Imani, Tajana Rosing\\
\IEEEauthorblockA{CSE Department, UC San Diego, La Jolla, CA 92093, USA\\
    \{bkhaleghi, sasalama, moimani, tajana\}@ucsd.edu}
}

\maketitle

\begin{abstract}
FPGA devices are continuously evolving to meet high computation and performance demand for emerging applications.
As a result, cutting edge FPGAs are not energy efficient as conventionally presumed to be, and therefore, aggressive power-saving techniques have become imperative.
The clock rate of an FPGA-mapped design is set based on worst-case conditions to ensure reliable operation under all circumstances.
This usually leaves a considerable timing margin that can be exploited to reduce power consumption by scaling voltage without lowering clock frequency.
There are hurdles for such opportunistic voltage scaling in FPGAs because (a) critical paths change with designs, making timing evaluation difficult as voltage changes, (b) each FPGA resource has particular power-delay trade-off with voltage, (c) data corruption of configuration cells and memory blocks further hampers voltage scaling.
In this paper, we propose a systematical approach to leverage the available thermal headroom of FPGA-mapped designs for power and energy improvement.
By comprehensively analyzing the timing and power consumption of FPGA building blocks under varying temperatures and voltages, we propose a thermal-aware voltage scaling flow that effectively utilizes the thermal margin to reduce power consumption without degrading performance.
We show the proposed flow can be employed for \emph{energy} optimization as well, whereby power consumption and delay are compromised to accomplish the tasks with minimum energy.
Lastly, we propose a simulation framework to be able to examine the efficiency of the proposed method for other applications that are inherently tolerant to a certain amount of error, granting further power saving opportunity.
Experimental results over a set of industrial benchmarks indicate up to 36\% power reduction with the same performance, and 66\% total energy saving when energy is the optimization target.

\end{abstract}

\IEEEpeerreviewmaketitle

\section{Introduction} \label{sec:intro}

The prevalence of computation-intensive workloads with high-performance requirements such as machine learning (ML) and data center applications \cite{putnam2014reconfigurable, lacey2016deep} accompanied with the advance of technology node have persuaded the FPGA vendors to integrate more resources with boosted clock rate in state-of-the-art FPGAs \cite{stratix10}.
This, together with slowed shrinking of FPGAs supply voltage \cite{ahmed2018automatic}, have pushed these devices to a point they consume power comparable to CPUs \cite{shen2019fast}.
Besides, there are prevailing energy-constrained applications in the {Internet of Things} (IoT) era, implemented in FPGAs, with the need for extreme energy efficiency \cite{blaauw2014iot, venkataramani2016efficient}.
All in all, more efficient power reduction approaches for FPGAs are now indispensable.

As FPGAs already employ a manifold of device-level optimization to throttle power consumption \cite{xilinx7}, more \textit{aggressive} power reduction techniques are gaining traction.
These techniques generally build upon the conservative timing margin ($d_g$) that is considered to compensate reliability threats in deep-nano technologies;
while an FPGA-mapped design is able to deliver an \textit{actual} clock period of $d_{V_{nom}}$ at nominal voltage $V_{nom}$, in practice, STA (static timing analysis) tools report an operating clock delay of $d_{V_{nom}} + d_{g}$ to make up for uncertainties such as voltage fluctuations, degradation, temperature, etc. \cite{shen2019fast}.
The aforementioned aggressive techniques exploit this available timing headroom to reduce the supply voltage of FPGAs down to $V_{low}$ for which the \textit{actual} clock period $d_{V_{low}}$ becomes equal or close to $d_{V_{nom}} + d_{g}$ (leaving no margin for guardbands).
Thus, the device still delivers the original performance at a lower voltage.
Note that \textit{aggressive} voltage scaling techniques are different from conventional DVFS (dynamic voltage and frequency scaling) that concurrently tunes the frequency and voltage based on per-task performance demand of applications.

Although lowering the supply voltage of processors and ASIC devices has been known to be an effective power saving technique \cite{bao2009line, lefurgy2011active, amrouch2018voltage}, there are limited studies presenting voltage scaling in FPGAs.
Voltage scaling (in particular,  aggressive and performance-aware one) in FPGAs is challenging mainly because:
\textbf{(a)} \emph{Critical Path} (CP) in an FPGA is design-dependent, which makes timing probing difficult under voltage scaling, especially considering the impact of temperature.
Therefore, in contrast to ASIC designs, a set of precalibrated stand-alone sensory circuits, e.g., ring oscillators and CP monitors \cite{lefurgy2011active} cannot accurately correlate the sensor frequency with all varying CPs.
\textbf{(b)} FPGA architectures are heterogeneous, comprising soft-fabric (i.e., programmable logic and routing resources), DSP cores, memory blocks (BRAM), etc.
These building blocks are tightly coupled, and each has a particular power/delay relation with supply voltage.
Considering the separate voltage rails provided for certain components, i.e., $V_{bram}$, $V_{core}$, and $V_{io}$ that can be regulated separately, finding efficient voltage points becomes design-dependent and challenging.
In other words, in multi-supply devices, multiple voltage combinations can lead to the target $d_{V_{nom}} + d_{g}$ boundary, while only one tuple yields minimum power.
This makes speculative voltage decrement no more efficient.
\textbf{(c)} Scaling the voltage of FPGAs is also constrained by the data corruption of configuration and memory SRAM cells.
In addition, as we will discuss in this paper, reducing the voltage of configuration cells unexpectedly {increases} FPGA power consumption in certain cases, calling for cautious analysis.

In this paper, we leverage the pessimistic thermal-induced timing slack to scale down FPGA operating voltage for power saving while tackling the aforementioned challenges as follows.\\
\textbf{(1)} We propose to incorporate thermal-aware voltage scaling in the FPGA design flow.
We first obtain the temperature-delay-voltage correlation of FPGA resources within the supported temperature range.
Then, we statically estimate the thermal distribution of applications to obtain the available timing headroom, for which voltages of different power rails can be efficiently determined based on the characterized library.
For further effectiveness, we also suggest online (i.e., dynamic) voltage adaptation based on the response of thermal sensors.
The proposed methods consider the voltage-temperature feedback loop and the separate power rails of specific resource types to yield maximum power efficiency.
\\
\textbf{(2)} We leverage the proposed flow of (1) to explore the \emph{energy} consumption, whereby we trade-off the performance and power consumption to achieve the minimum energy point.
This is desirable for a majority of edge and IoT applications for which total energy consumption is the utmost concern.
\\
\textbf{(3)} Our approaches mentioned in (1) and (2) are deterministic, as they guarantee timing closure.
Nonetheless, many use-cases such as image processing and machine learning applications can tolerate a certain level of computation errors, which gives an opportunity for voltage \emph{over-scaling}.
However, it needs an examination of applications under these non-ideal conditions to get a glimpse of produced error in the output.
We propose a primary FPGA simulation framework to be able to evaluate an FPGA-mapped design under voltage-scaling, i.e., when the delay of resources varies.
Using the proposed framework, we map demonstrative machine learning applications into FPGA fabric and examine the impact of voltage \emph{over-scaling} on extra power gain versus accuracy drop.

\section{Background and Related Work}\label{sec:rel}


\subsection{FPGA Architecture} \label{subsec:arch}

\begin{figure}[t]
  \centering
  \includegraphics[width=0.5\textwidth]{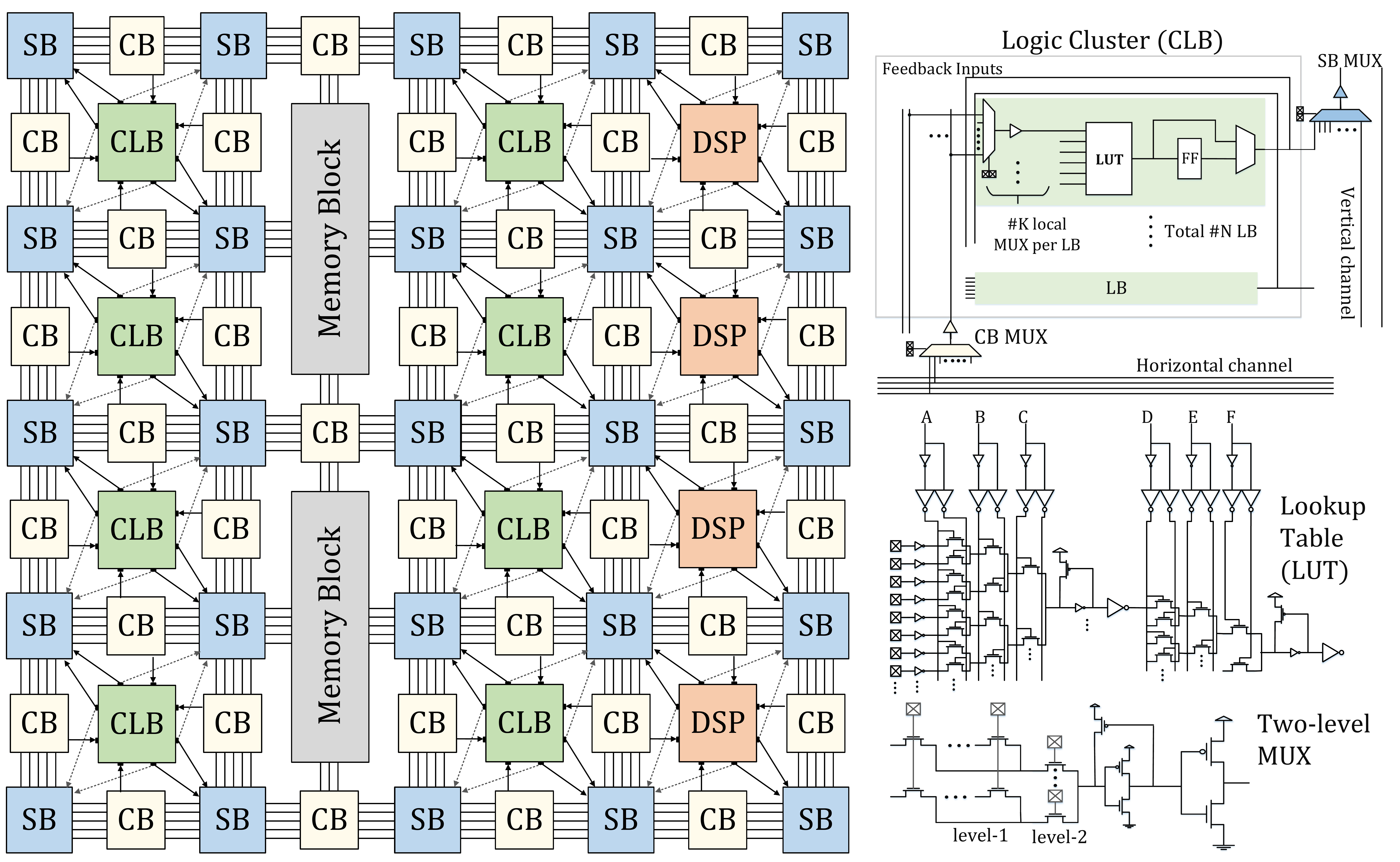}
  \caption{Tile-based FPGA architecture (left), and constructing building blocks (right) \cite{chiasson2013should}.}
  \label{fig:arch}
\end{figure}

Fig. \ref{fig:arch} illustrates the architecture of conventional tile-based FPGAs.
The architecture comprises tiles of logic clusters (a.k.a CLBs or slices) that bind together using the configurable switch boxes (SBs) and connection blocks (CBs) to implement larger functions.
Each logic CLB consists of  $N$ (e.g., $N=10$) $K$-input look-up tables (LUTs) each of which is capable to implement Boolean expressions up to $K$ variables.
SB multiplexers are located in the intersection of horizontal and vertical channels (wire tracks) to enable connectivity and bending of nets.
These multiplexers are also responsible for connecting outputs of logic resources (LUTs, FFs, carry chain, etc.) to global routing, i.e., to horizontal and vertical channels.
Analogously, CB multiplexers pass the selected global wires into the logic clusters.
Specific FPGA columns are repetitively dedicated to Block RAMs and DSP cores.
The majority of logic and routing resources have a multiplexer-like structure, mainly implemented as two-stage multiplexers that have been shown to provide optimal area-delay efficiency \cite{lewis2005stratix}.
These resources often drive large loads, especially the global routing resources that have a high number of long fanout wires, so are augmented with large output buffers to improve performance.

\subsection{Related Work} \label{subsec:related}

While many previous studies have attempted to reduce FPGA power dissipation by power-gating, conventional DVFS, configurable dual supply voltage, control of signal levels, architectural innovations, etc. \cite{seifoori2018introduction}, there is a limited number of works that leverage the available timing margin to reduce voltage without scarifying performance. 
Authors of \cite{levine2014dynamic} explore the timing headroom of FPGA-mapped designs by gradually
reducing the voltage (keeping the frequency fixed) until observing the error.
They detect the error by inserting a shadow register per each CP with a phase-shifted clock to detect the data mismatch caused by voltage reduction.
As there are a huge number of CP and near-CP paths in a large synthesis-flattened design, the area and power overhead of this technique can be excessive. Also, measuring the slack in this technique depends on capturing the signal traversing the CP, while the CP may not be controllable at the runtime.
Furthermore, the introduced capture registers cannot be used for paths that head to hard blocks such as BRAM.
The study in \cite{nunez2017adaptive} addresses the latter issue by replicating such paths and inserting an end-point proxy register (instead of the hard block) where an error-detector circuitry checks timing violations. However, the error-detector itself imposes delay, so a timing error raised in the original CP might propagate into the memory before being detected.

In \cite{zhao2018robust}, the authors propose a two-step self-calibrating voltage scaling scheme by exploiting the available timing slack of thermal margin.
CPs of a design are extracted by using the STA tool and are then implemented on the FPGA fabric. Afterward, for different values of temperature and core voltage ($T$ and $V_{core}$), the maximum frequency is obtained by gradually increasing it until error is observed by the implemented error-detection circuit.
This approach does not consider the thermal distribution within the chip \cite{amouri2013accurate} where a CP may experience varied temperature in reality.
This either results in timing violation by ignoring the parts of CP that reside in hot (hence, slower) tiles or gives non-optimal results if a related thermal margin is considered.
In addition, the STA tool reports the CPs according to worst-case condition while CPs might change at lower temperatures.
Thus, a larger number of near-CP paths need to evaluate which makes the entire process further cumbersome.
BRAM and soft-logic voltage rails also are not separately considered, so the minimum employed voltage will be limited by the one that violates the timing first.

Finally, recent work in \cite{salami2018bram} examines the voltage scaling of FPGA BRAMs.
The authors showed that up to 39\% of BRAM voltage can be reduced without observing any error.
Though it is promising in reducing the power of BRAMs by one order of magnitude, a trial-and-error based approach does not guarantee correct functionality as it is infeasible to examine all the inputs. Moreover, the overall efficiency is limited to the power of BRAMs.

Aforementioned techniques are all speculative as they decrease the voltage until observing an error.
This overlooks errors that emerge gradually due to violating the guardbands (e.g., timing error as a result of degradation \cite{khaleghi2019estimating}) or may arise abruptly due to voltage transients \cite{shen2019fast}.
Recently, work in~\cite{shen2019fast} showed a margin of over 36\% is needed for voltage transients as a result of load transients, and this margin is already considered in STA tools.
Nevertheless, as voltage transients occur infrequently, the speculative voltage scaling methods do not take them into consideration, which will lead to timing violation at certain conditions.
Our voltage scaling approach is different from previous studies as we incorporate it in the FPGA design flow by characterizing the resources during FPGA architecting.
This eliminates the arduous task of voltage-timing speculation and guarantees timing.
Our method precisely considers the correlation of temperature, delay, voltage, and power of resources and separate power rails of soft-fabric and memories, so it yields maximum efficiency by setting the optimal core and BRAM voltages as well as accurately estimating the timing according to the thermal distribution of the blocks.
Finally, \emph{over-scaling} of voltages needs timing simulation to observe the impact of timing violations.
We enable it by our novel FPGA simulation flow.
It is noteworthy that, similar to our work, the studies in \cite{khaleghi2019thermal} and \cite{salamat2019workload} also propose \textit{deterministic} frequency and/or voltage adaption through resource characterization.
Nonetheless, \cite{khaleghi2019thermal} targets performance boosting by using non worst-case guardband in lower temperatures, and \cite{salamat2019workload} adjusts both the frequency and voltage in non-peak workloads to save power (hence, has nothing to do with the temperature).



\section{Proposed Method} \label{sec:prop}

\subsection{Preliminary}\label{subsec:preliminary}
FPGA flow is different from conventional standard-cell based design of ASICs.
Thus, we first elaborate the setup of experiments used in the rest of the paper before detailing the proposed method.

\begin{figure*}
\begin{center}

\subfloat[(Different) delay-temperature relations.\label{fig:D-T}]{
                \includegraphics[width=0.32\textwidth, height=2.6cm]{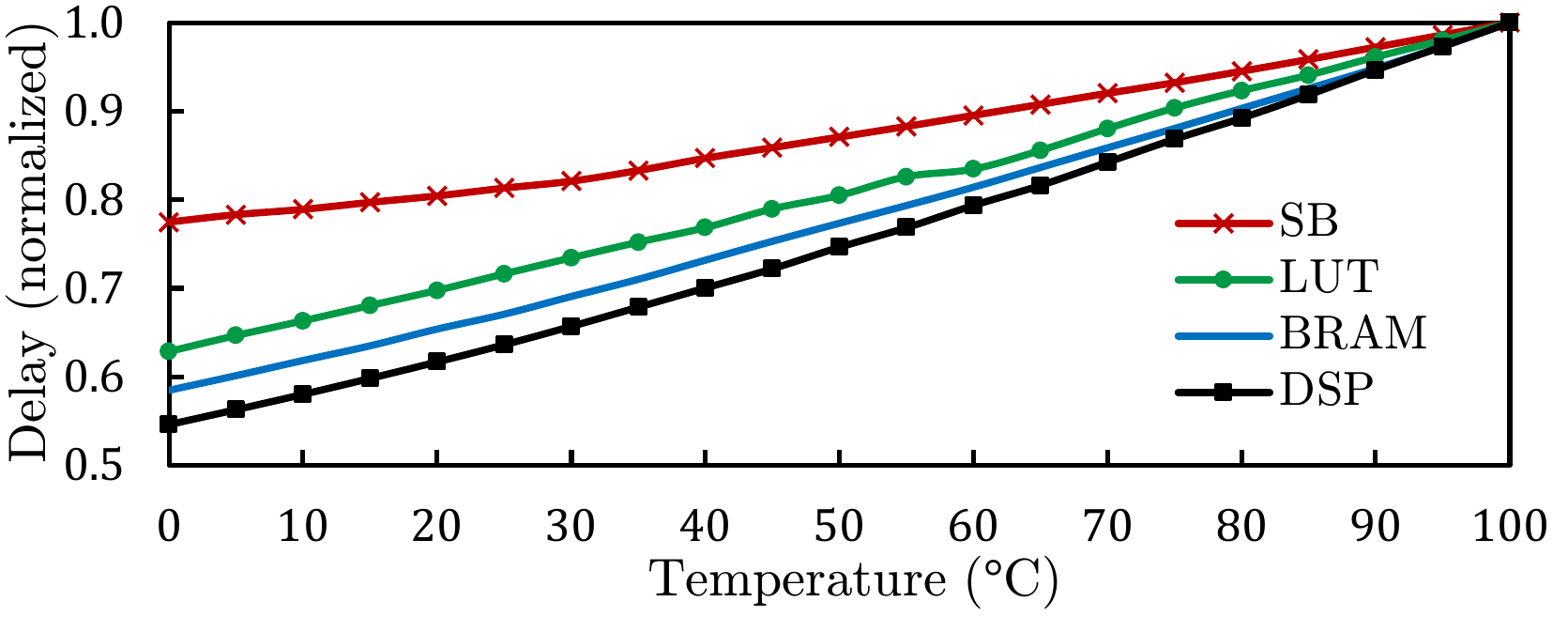}
}
\hspace{-0.3cm}
\subfloat[(Different) delay-voltage relations of resources.\label{fig:D-V}]{
                \includegraphics[width=0.32\textwidth, height=2.6cm]{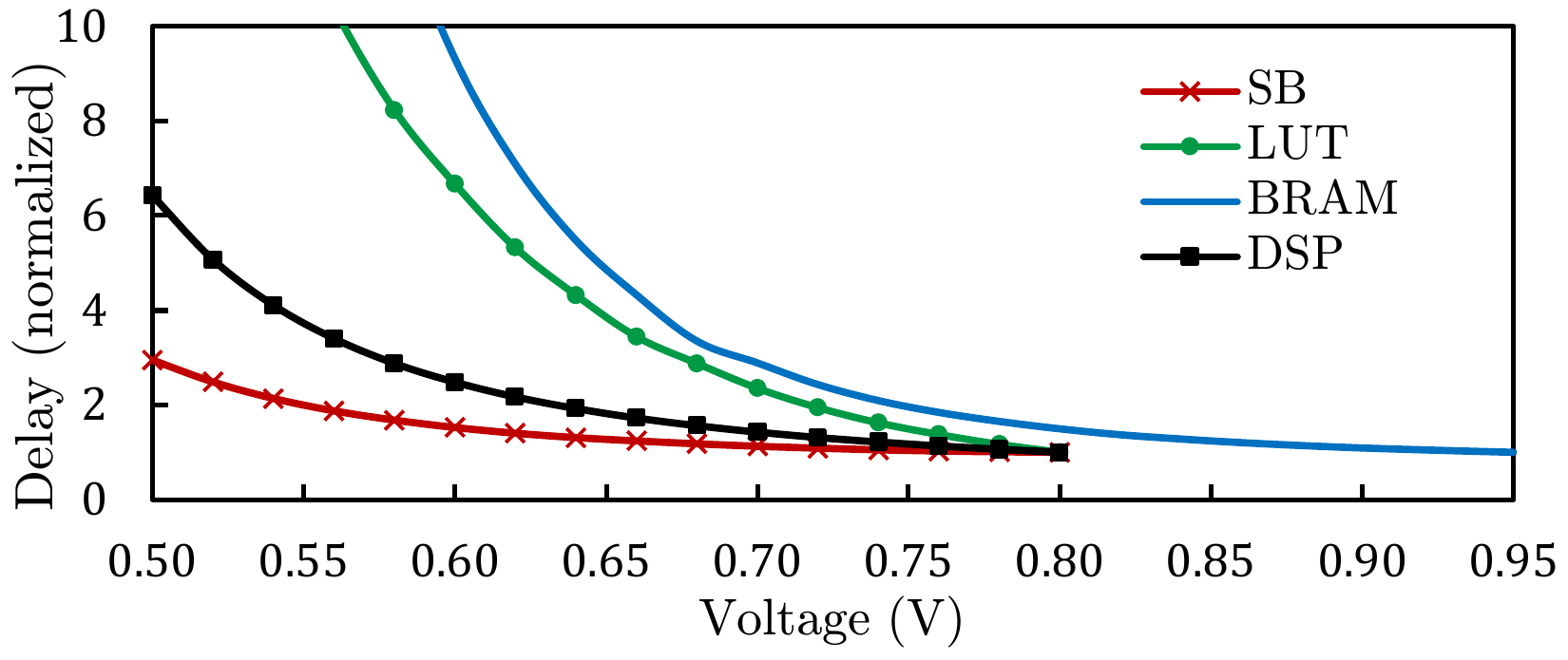}
}
\hspace{-0.3cm}
\subfloat[(Different) power-voltage relations of resources.\label{fig:P-V}]{
                \includegraphics[width=0.32\textwidth, height=2.6cm]{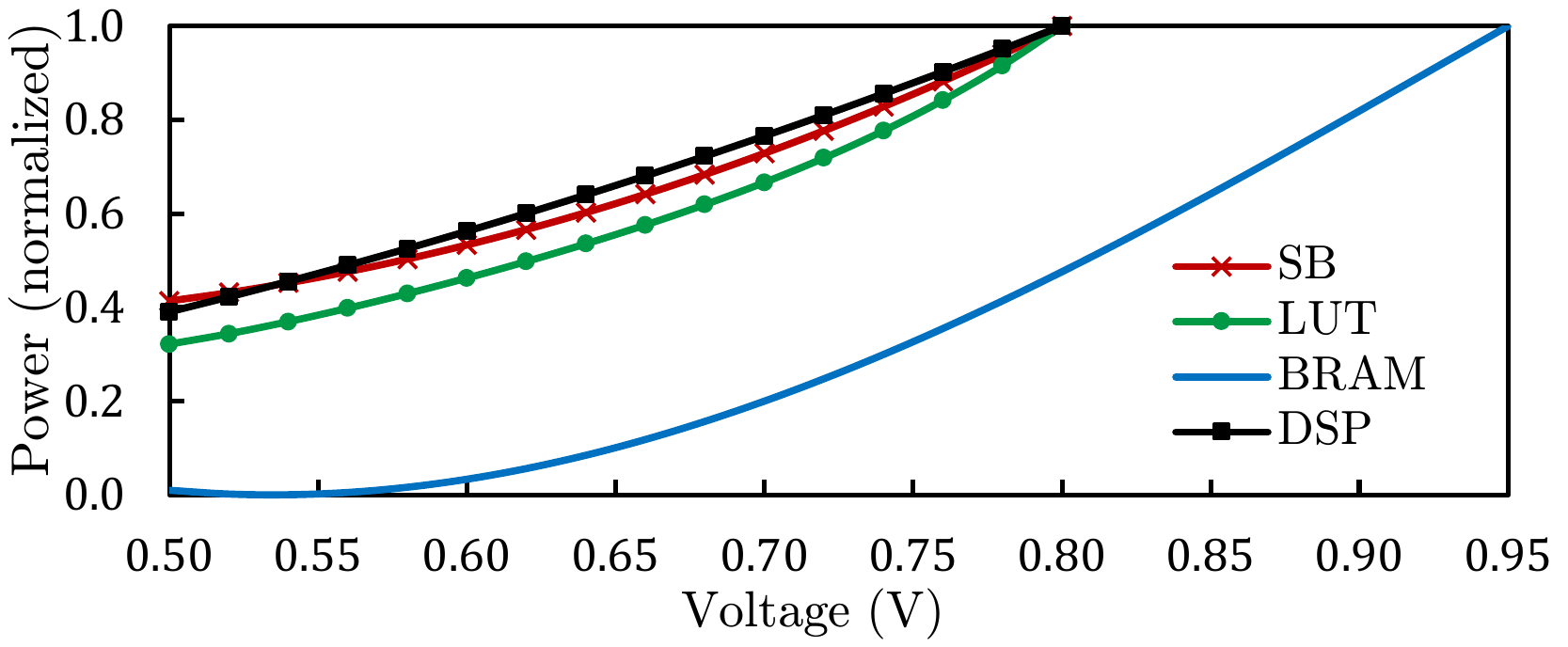}
}
\caption{Behavior of different FPGA resources under varying temperature and voltages.\label{fig:motivation}}
\vspace{-0.5cm}
\end{center}
\end{figure*}

\textbf{$\bullet$ Power and Delay.}
We use circuit-level simulations to obtain the delay and power of FPGA resources.
For this end, we use the latest version of COFFE \cite{chiasson2013coffe} that generates and characterizes an accurate netlist of FPGA resources according to the given architectural description using comprehensive circuit-level HSPICE simulations.
Besides the description of the target FPGA architecture, COFFE also requires technology process of transistors, for which we use 22\,nm predictive technology model (PTM) \cite{ptm}.
COFFE also generates and evaluates the memory blocks of FPGA and has been shown to have a suitable delay and exact area match with commercial FPGAs \cite{yazdanshenas2017don}.
Similar to configuration SRAM cells, the core of the memory blocks (i.e., eight-transistor dual-port SRAM cells) is implemented by 22\,nm high-threshold low-power transistors \cite{tuan200690nm}, which throttles their leakage power by two orders of magnitude.
As we will show in the rest of this section, our simulations show a similar power trend to commercial FPGAs.

COFFE does not model DSP blocks.
We thus develop the DSP HDL code based on the Stratix IV description \cite{stratixiv} and characterize it using Synopsys Design Compiler with NanGate 45\,nm open cell library \cite{nangate2008california}.
Then we scale the results to 22\,nm based on measuring scaling factors of a selected set of cells at 45\,nm and 22\,nm technologies.
Based on PrimeTime report, the developed DSP consumes 4.6\,mW at 250\,MHz, which is comparable to a 28\,nm DSP that dissipates 5.6\,mW at the same frequency \cite{xpower}.
To characterize the delay and power of programmable resources at different temperature and voltages, we sweep the parameters of COFFE-generated netlists in HSPICE simulations.
For DSP, we create a set of standard-cell libraries using NanGate netlists by the means of Synopsys SiliconSmart so we could characterize the DSP at different operating conditions.

\begin{table}[t]
\centering
\caption{FPGA architecture parameters used in COFFE}
\label{tab:coffe}
\resizebox{0.46\textwidth}{!}{
\begin{tabular}{lclc}
\hline
Parameter & Value & Parameter & Value \\ \hline
K               & 6            & $SB_{mux}$ size      & 12              \\ 
N               & 10           & $CB_{mux}$ size       & 64            \\ 
Channel tracks     & 240          & $local_{mux}$ size     & 25      \\ 
Wire segment length    & 4     & $V_{core}, V_{bram}$  & 0.8\,V, 0.95\,V     \\ 
Cluster global inputs       & 40           & BRAM            & 1024 $\times$ 32\,bit              \\ \hline
\end{tabular}
}
\end{table}

\textbf{$\bullet$ FPGA Flow.}
We use VTR 7.0 (Verilog-to-Routing) \cite{luu2014vtr} toolchain that enables defining a customized FPGA architecture and place and route the benchmarks.
We select our benchmarks from VTR repository that belong to a wide variety of applications (vision, math, communication, etc.), contain single- and/or dual-port memory blocks as well as DSP blocks, with an average of over 23,800 $6$-input LUTs (maximum over 106\,K).
We use FPGA architecture parameters similar to Intel Stratix devices \cite{lewis2009architectural, chiasson2013should} in COFFE and VPR\footnote{VPR (Versatile Place and Route) is the P\&R tool in VTR toolchain.} placement and routing, which is summarized in Table \ref{tab:coffe}.
COFFE uses these parameters to generate SPICE netlist and area and delay report, which are then fed into VTR to place and route the benchmarks.
$K$ and $N$ are the sizes of LUTs and number of logic blocks in a cluster (see Fig. \ref{fig:arch}), which are chosen to be 6 and 10 in accordance with Intel devices.
Estimating the power consumption of applications (for both thermal simulation and power saving estimation) needs also their signal activity, for which we use ACE 2.0 \cite{lamoureux2006activity}.
We modified VPR to enable timing analysis at different scenarios using the characterized libraries.

\textbf{$\bullet$ Thermal Simulation.}
We use HotSpot 6.0 \cite{zhang2015hotspot} for thermal simulations.
Inputs of HotSpot are device floorplan, power trace (or average power values), and device configuration parameters.
For the floorplan file, we divide the device floorplan into a two-dimensional array of FPGA tiles with the areas reported by COFFE. We assume CLB tiles are square, and the heights of DSP and memory blocks are $4\times$ and $6\times$ of CLB tiles \cite{luu2014vtr}.
The number of tiles and location of each tile can be obtained from the placement and routing outputs reported by VPR.
Leakage and dynamic power of each tile is obtained based on the current temperature and activities of resources of the tile.
Finally, for the HotSpot configuration file, we change the parameters according to validated FPGA parameters in \cite{velusamy2005monitoring}.
We adjust the convective resistance to concur with an effective thermal resistance ($\theta_{JA}$) of contemporary FPGAs, i.e., we tune $\mathtt{r\_convec}$ such that when the total power of given power trace is set to 1 Watt, the reported temperature by HotSpot equals $\theta_{JA}$.
We examine the efficiency of the proposed technique by using a typical $\theta_{JA}$ of $2\,\sfrac{\degree C}{W}$ as in today's Intel and Xilinx devices (e.g., Virtex--7 and Stratix V) \cite{xpower,intelpower}, and a pessimistic thermal resistance of $12\, \sfrac{\degree C}{W}$, corresponding to their mid-size devices (such as Spartan--7 or Artix--7) with still airflow.

\subsection{Proposed Thermal-Aware Voltage Scaling Flow}\label{subsec:flow}

\textbf{$\bullet$ Motivation.} Fig. \ref{fig:motivation} gives perception on how temperature margin can be leveraged for power reduction.
This figure is obtained using the experimental setup explained in the previous subsection.
Numbers were not in the same range, so we normalized each one to its base value at $100\degree$C and $0.8$V for the sake of clear illustration.
According to Fig. \ref{fig:motivation}(a), although FPGA timing analysis reports the worst-case to ensure timing meets in all scenarios \cite{timing}, in practice, resources have a smaller delay at lower temperatures.
For instance, at 40\,$\degree$C, delay of switch box (\textcolor{Red}{$\mathrel{\ooalign{\hss $\times $\hss\cr ---}}$} SB) is 0.85$\times$ of its delay at worst-case temperature\footnote{We assume an upper-bound of 100\,$\degree$C for junction temperature \cite{stratixiv}.}.
This gap can be utilized for voltage reduction.
Based on Fig. \ref{fig:motivation}(b), 0.68\,V is the point wherein this margin is fully utilized, i.e., delay of 40\,$\degree$C increases by $\frac{1}{0.85\times}$ and becomes equal to delay at worst-case temperature.
Eventually, Fig. \ref{fig:motivation}(c) reveals this 120\,mV reduction of voltage shrinks the switch box power down by 32\%.
As mentioned before, memory block comprises of low-threshold transistors, so uses a different power rail with a voltage higher than datapath (core) transistors.
Other non-memory resources show a {\raise.17ex\hbox{$\scriptstyle\mathtt{\sim}$}}$V^2$ relation with voltage while BRAM observes a more dramatic power reduction as voltage scales.
As is evident from the figure, different resources exhibit different delay behavior as temperature and voltage change.
This stems from different sizing of transistors, input slope, output capacitance, etc., as detailed in previous works \cite{amrouch2017optimizing, amrouch2018voltage}.

We can get several insights from Fig. \ref{fig:motivation} and above discussion.\\
\textbf{(a)} Replica circuits such as ring oscillators used to correlate the temperature/voltage with frequency of ASICs are not a viable solution to track timing because, in FPGAs, CPs are design-dependent and made from different types and count of resources.
Apparent from Fig. \ref{fig:motivation}(a) and (b), designs bounded by routing (SB) have a totally different performance behavior compared to logic (LUT) bounded ones when temperature or voltage changes, so a set of representative paths fails to resemble all paths accurately.\\
\textbf{(b)} Previous studies rely on worst-case reported paths to check the timing (or to insert error detectors) while lowering the voltage.
Nonetheless, from Fig. \ref{fig:motivation}(a) and (b) we can infer a non-CP path may become CP at lower temperature or voltage.
For instance, LUT delay severely increases at lower voltages, so the delay of LUT-bounded paths can exceed originally reported SB-bounded paths.
This signifies cautious timing analysis in voltage scaling.\\
\textbf{(c)} More importantly, even if timing analysis of an FPGA-based design were possible under arbitrary (T, V) pairs, efficient voltage scaling would be still challenging because, as shown in Fig. \ref{fig:motivation}(c), resources enjoy differently from voltage reduction.
For instance, memory block shows better power saving as voltage scales, while, on the contrary, its delay also increases more under voltage scaling.
Thus, for a certain timing headroom, there is a trade-off between power gain and increase of delay (i.e., use up of margin) when there are multiple power rails.
This justifies why temperature-voltage-delay correlation and voltage-power libraries are indispensable for a reliable and efficient thermal-aware voltage scaling.

\textbf{$\bullet$ Thermal-Aware Voltage Scaling Flow.}
Taking the above-mentioned insights into consideration, in the following we present our voltage scaling algorithm as the core of our energy efficiency technique, and then further elaborate it by exemplifying case-studies.
Algorithm \ref{algorithm1} can be either simply integrated into the current FPGA flow stack (i.e., in the original timing analysis step), or attached as an additional step.
In either case, it relies on a pre-characterized library of delay and power, as detailed in Section \ref{subsec:preliminary}.
If Algorithm \ref{algorithm1} is used as an additional step, then it needs to get the post place and route netlist.
Other inputs of the algorithm are maximum temperature surrounding the FPGA board, and sample inputs or activities (we further discuss it later).

\begin{algorithm}[t]
\small
\KwIn{$netlist$: Placed and routed design}
\KwIn{$T_{amb}$: Ambient temperature}
\KwIn{$\overrightarrow{\alpha}$: Input activities / sample inputs}

$\overrightarrow{T}_{m \times n} = [T_{amb}, \cdots, T_{amb}]$ // $m,n: $ FPGA grid size \\
$\overrightarrow{\Delta T}_{m \times n} = [\infty, \cdots, \infty]$ \\
$d_{worst} = \mathcal{T}(netlist, {T}_{max}, V_{core_{max}}, V_{bram_{max}})$\\

\While{$\| \overrightarrow{\Delta T} \|_{\infty} > \delta_T$}{
$\underset{V_{core},\ V_{bram}}{\text{min}}\overrightarrow{P}_{lkg}(\overrightarrow{T}, V_{core}, V_{bram}) \ + $\\
$\qquad \qquad \ \ \overrightarrow{P}_{dyn}(netlist, \overrightarrow{\alpha}, f_{worst}, V_{core}, V_{bram})$\\
$\qquad \text{s.t.} \ \mathcal{T}(netlist, \overrightarrow{T}, V_{core}, V_{bram}) \leq d_{worst}$

$\overrightarrow{T}_{old} = \overrightarrow{T}$\\
$\overrightarrow{T} = HotSpot(\overrightarrow{P}_{lkg} + \overrightarrow{P}_{dyn})$ \\
$\overrightarrow{\Delta T} = \overrightarrow{T} - \overrightarrow{T}_{old}$\\
}
\Return $V_{core}, V_{bram}$
\caption{{\sc} Thermal-Aware Voltage Selection} \label{algorithm1}
\end{algorithm}
\normalsize

FPGA thermal estimation usually relies on a single total power and ambient temperature value to estimate the junction temperature \cite{xpower}, however, we divide the target FPGA into a grid of $m$\,$\times$\,$n$ tiles, for $m$ and $n$ being the number of FPGA rows and columns.
It improves the accuracy of thermal estimation and helps to catch potential hotspot regions, where the blocks have higher delay than the rest of the board, hence need fine-grained timing analysis for both accuracy and efficiency (i.e., to avoid under- or over-estimation of timing).
$d_{worst}$ is the delay that conventional one-size-fits-all timing analysis $\mathcal{T}$ of FPGA reports under nominal memory and core voltages and maximum temperature \cite{timing} while also considers some margin for reliability issues such as voltage transients \cite{shen2019fast}.
Thus $d_{worst}$ is the target delay that our algorithm attempts to deliver with lower voltages.
One drawback of previous voltage scaling approaches \cite{levine2014dynamic, zhao2018robust} is they invade this reliability margin when they speculatively reduce the voltage until observing an error in the output, as the error does not show up in regular conditions.
The core of the algorithm is a loop where, based on previously obtained temperature for each tile (set to $T_{amb}$ initially), it finds the ($V_{core}, V_{bram}$) pair that minimizes the power while watches over the delay of candidate pair to not exceed $d_{worst}$.
In the first iteration of the algorithm, it explores all $|V_{core}| \times |V_{bram}|$ pairs.
In the next iterations, execution time can be significantly reduced by limiting the search to the boundaries of the previous solution, making subsequent iterations $O(1)$.
Note that in both timing analysis and power calculation (lines 5--7), each tile has its own activity and potentially different temperature, so we use vectors of length $m$\,$\times$\,$n$ to store the values associated with each tile.
Temperature affects the leakage power and delay of the tile resources, while activity affects the dynamic power. Hence, $\overrightarrow{T}$ is passed to $\overrightarrow{P}_{lkg}$ calculation and timing ($\mathcal{T}$) analysis, while $\overrightarrow{\alpha}$ is passed to $\overrightarrow{P}_{dyn}$ to estimate dynamic power at $d_{worst}$ (clock cycle will be always $d_{worst}$).
Finally, the power values are imported in a thermal simulator to update temperatures of tiles.
This procedure repeats until reaching a steady-state temperature.
For thermal simulation in our experiments we use HotSpot 6.0 \cite{zhang2015hotspot}, with setup already detailed in Section \ref{subsec:preliminary}.

\begin{figure}[t]
  \centering
  \includegraphics[width=0.42\textwidth]{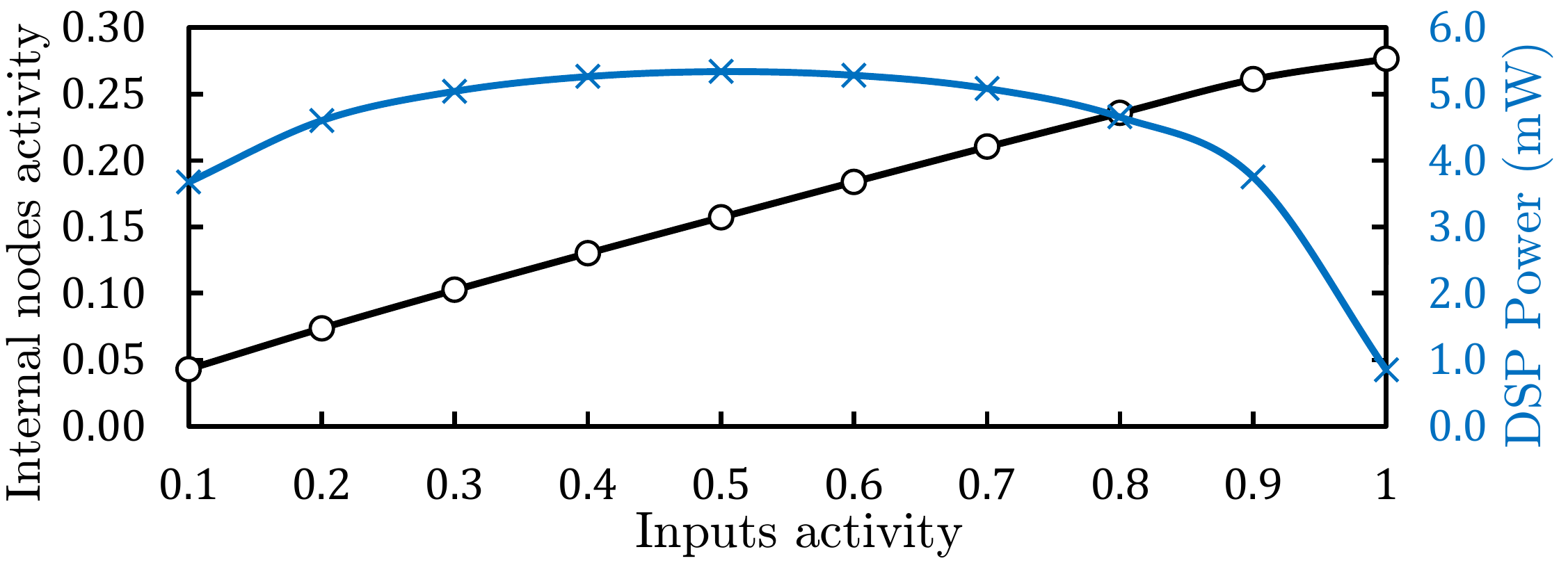}
  \caption{Activity of benchmarks nodes for different activities of primary inputs (left/blue), and DSP power at different activities of its inputs (right/red).}
  \label{fig:activity}
\end{figure}

\textbf{$\bullet$ Static and Dynamic Implementations.}
Static implementation of the proposed technique is straightforward as both core and memory voltages are determined during the configuration of design, whether by incorporating Algorithm \ref{algorithm1} in original timing analysis of FPGA, or using it as a post-routing addendum.
As the voltages remain fixed in the field operation, the algorithm needs to consider the corner case of the programmed design, i.e., the highest temperature it might reach.
A noteworthy point here is that activities of internal nodes of a design do not linearly correspond to activities of primary inputs.
In Fig. \ref{fig:activity}, when signal activity factor ($\alpha$) of benchmarks inputs increases from 0.1 to 1, activity of internal nodes (averaged over all 10 benchmarks) increases from 0.05 to {\raise.17ex\hbox{$\scriptstyle\mathtt{\sim}$}}0.27, which is significantly less than $\alpha=1$ considered for primary inputs.
In addition, in some blocks such as DSP, the increase in activity of primary inputs does not necessarily translate to increase of power.
As can be seen from Fig. \ref{fig:activity}, DSP power increases by only around 37\% when its inputs activities raise from 0.1 to 0.3, then its power saturates until $\alpha \in$ [0.3, 0.7], and declines thereafter.
This behavior of power is because the frequently changing inputs offset each other more often (e.g., when both inputs of an XOR function change in a clock, its output remains the same).
All in all, by caring for the worst-case input activity, the proposed static scheme guarantees reliable operation at corners without overly pessimistic activity estimation, though the ambient temperature needs to consider maximum possible.

\begin{figure*}
\begin{center}

\subfloat[Voltage of core and memory power rails.\label{fig:mkDelayWorker-V}]{
                \includegraphics[width=0.32\textwidth, height=2.6cm]{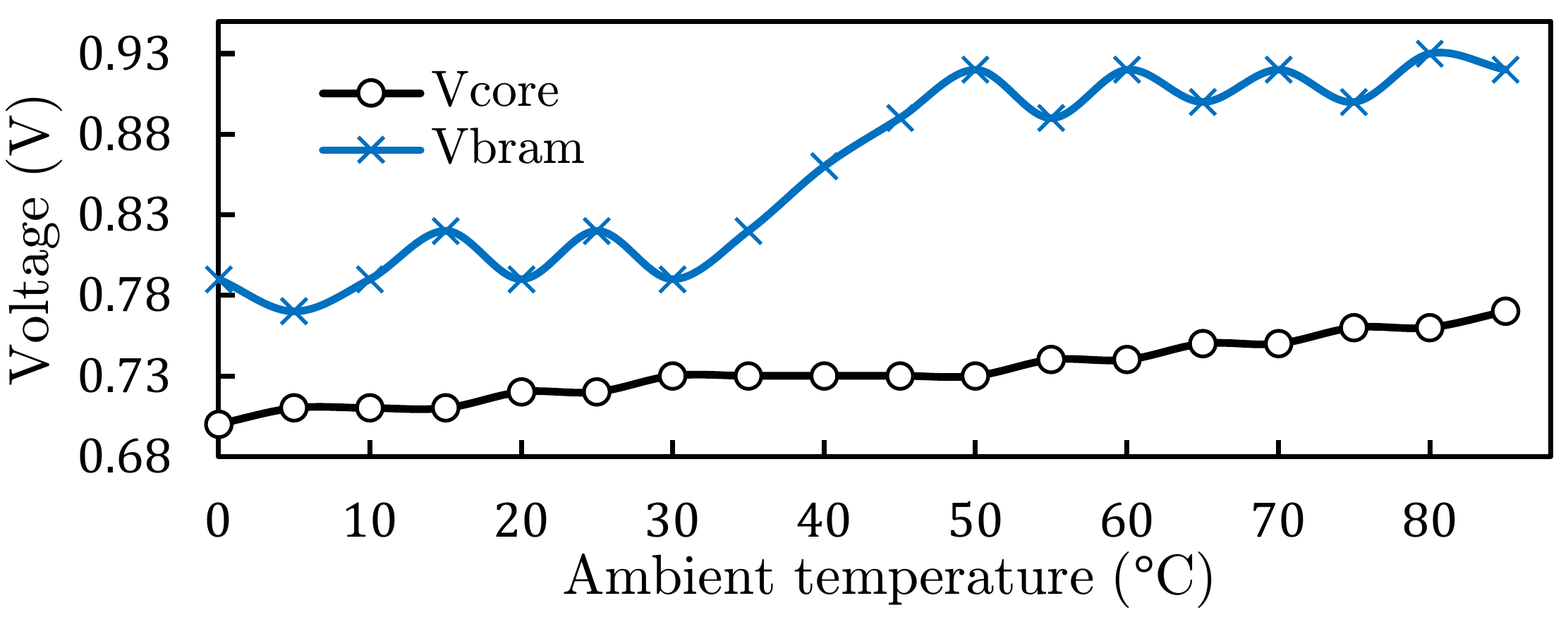}
}
\hspace{-0.3cm}
\subfloat[Total consumed power bounds.\label{fig:mkDelayWorker-P}]{
                \includegraphics[width=0.32\textwidth, height=2.6cm]{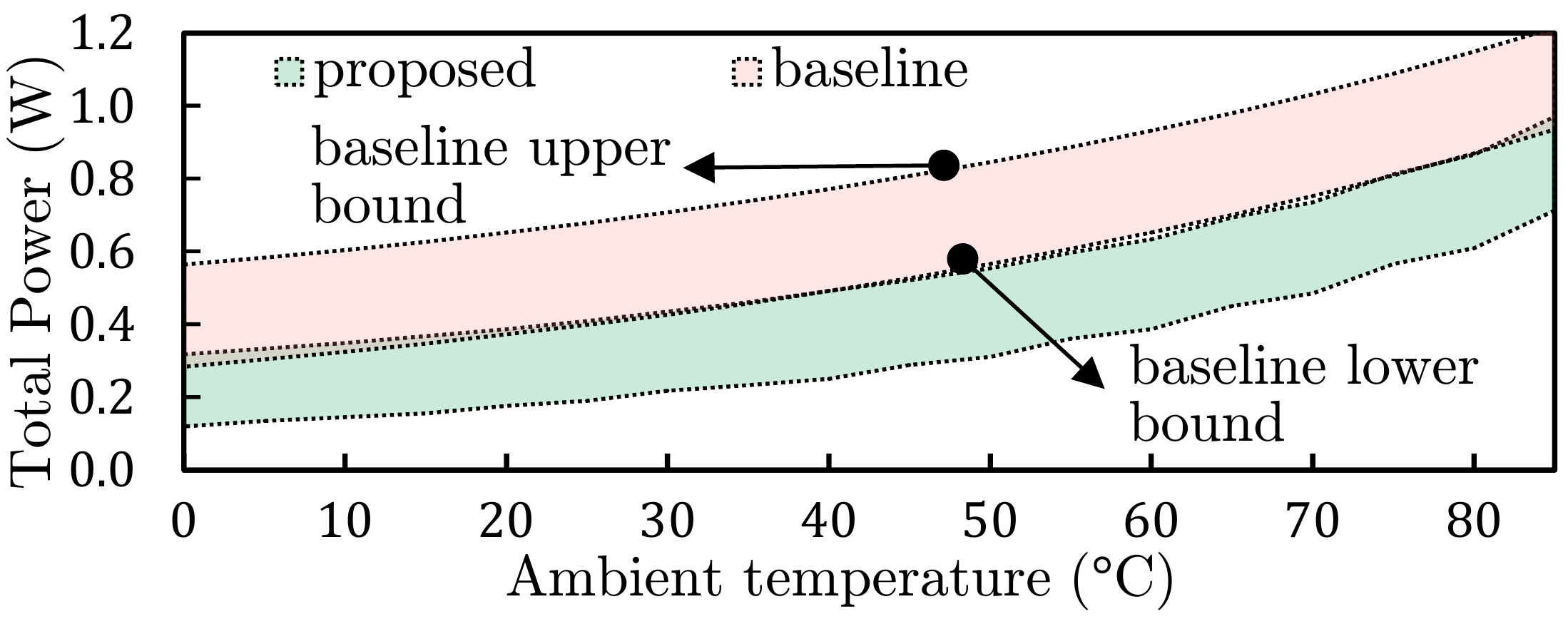}
}
\hspace{-0.3cm}
\subfloat[Increase of junction temperature bounds.\label{fig:mkDelayWorker-T}]{
                \includegraphics[width=0.32\textwidth, height=2.6cm]{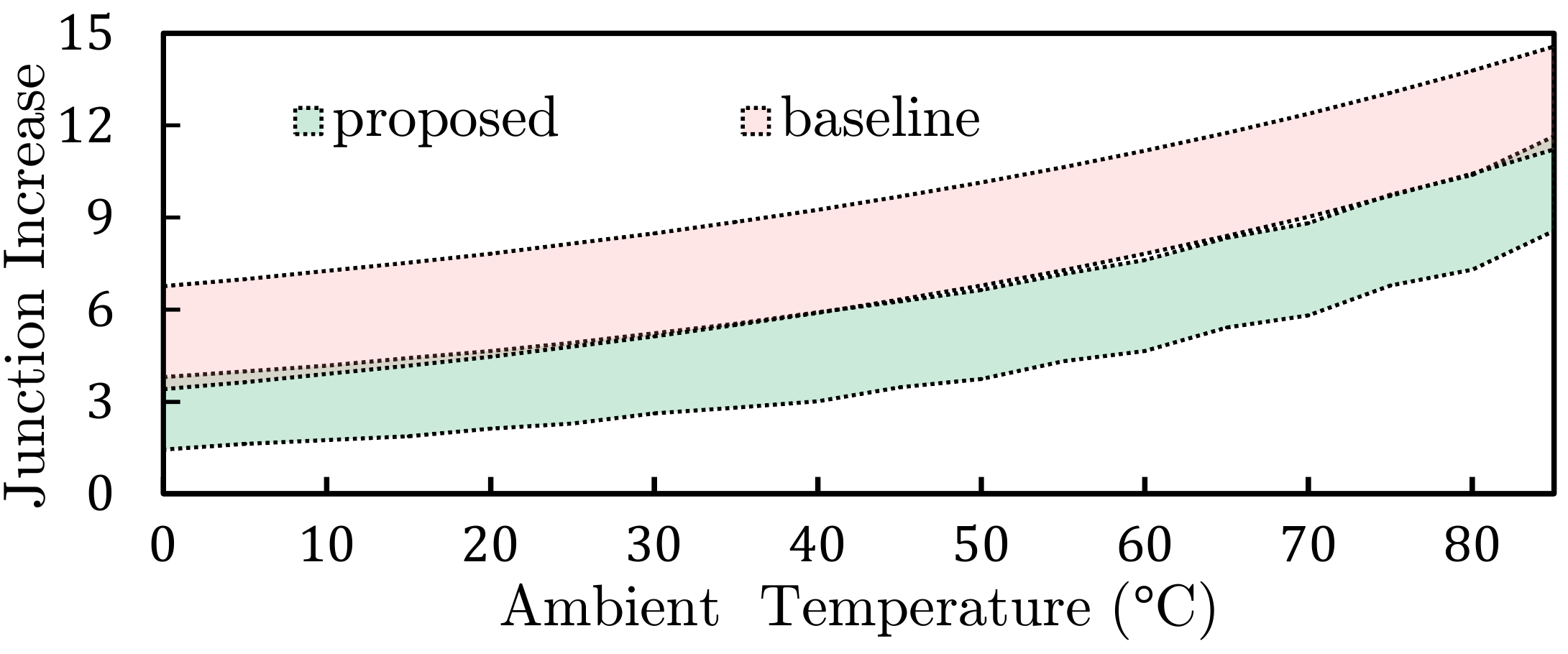}
}
\caption{Outputs of Algorithm \ref{algorithm1} for $\mathtt{mkDelayWorker}$ benchmark under different ambient temperaturs.\label{fig:mkDelayWorker}}
\vspace{-0.5cm}
\end{center}
\end{figure*}

The proposed thermal-aware voltage scaling scheme can be implemented online (dynamic) to avoid pessimistic assumption on the temperature bound.
Instead of thermal simulation, online scheme reads the junction temperature using on-board sensors available on all contemporary FPGAs.
For instance, Intel devices already contain temperature sensing diodes (TSD) with instantiatable IP cores having their internal clock source that can output the junction temperature with a resolution of 10 bits in 1,024 clock cycles (i.e., 1\,ms) \cite{intelsensor}.
Therefore, during the configuration of each design, we create a look-up table with temperature $T$ as its keys and $(V_{core}, V_{bram})$ as the values that minimize the power for that $T$.
Optimal $(V_{core}, V_{bram})$ of each temperature can be obtained in the same way explained above for the static approach.
Reading the temperature with steps of few milliseconds is large-enough to allow on-chip voltage regulators (such as Intel on-the-fly regulators) to adjust the voltage \cite{burton2014fivr}, and yet is small enough to avoid temporal heat-up mismatch that takes orders of seconds \cite{tian2019temporal}.
The sensed junction temperature can be directly used as VID (voltage identification) for the programmable integrated voltage regulator to adjust the voltage with the pre-loaded values \cite{burton2014fivr}.
A thermal margin (e.g., 5\,$\degree$C) might be considered to account for the error of TSDs and potential spatial thermal gradients \cite{ebi2011economic}.

\textbf{$\bullet$ Case-study.}
To elaborate our method, we use $\mathtt{mkDelayWorker}$ benchmark with 6,128 LUTs and 164 memory blocks that VPR mapped it to a 92$\times$92 grid device due to its high BRAM demand, with a frequency of 71.6\,MHz.
Our simulations show the device consumes a leakage power of 0.367\,W at 25\,$\degree$C (considering all used and unused resources), while the closest Intel device (Stratix V 5SGSD3) is 1.5$\times$ in size with a power of 0.646\,W.
This 1.76$\times$ power ratio is acceptable considering the size difference and more advanced technology we use in our simulations (22\,nm versus 28\,nm).

Fig. \ref{fig:mkDelayWorker} shows the results of our static voltage scaling scheme on $\mathtt{mkDelayWorker}$ benchmark.
We assume ambient (near-board) temperature range from 0\,$\degree$C up to 85\,$\degree$C as previous studies have shown that board temperature of datacenter FPGAs can reach up to {\raise.17ex\hbox{$\scriptstyle\mathtt{\sim}$}}70\,$\degree$C \cite{putnam2014reconfigurable}.
Fig. \ref{fig:mkDelayWorker}(a) shows that, moving from 0\,$\degree$C to 85\,$\degree$C, generally both $V_{core}$ and $V_{bram}$ increase towards their nominal 0.8\,V and 0.95\,V values to meet timing in worst-case junction temperature.
Small fluctuations of BRAM voltage at certain points is to yield maximum power saving.
For instance, at 30\,$\degree$C, $V_{core}$, $V_{bram}=$\,(0.73,\,\textbf{0.79}) while at 25\,$\degree$C, $V_{core}$, $V_{bram}=$\,(0.72,\,\textbf{0.82}), however, we expected BRAM voltage to be lower for 25\,$\degree$C.
This is because actually the 10\,mV reduction of $V_{core}$ at 25\,$\degree$C is worth the 30\,mV increase of $V_{bram}$; as we examined, it resulted in a power of 410\,mW while experiments showed that the other combination ($0.73$,\,$0.79$) would consume 420\,mW ($>$\,410\,mW).
It indicates preciseness of our technique in determining most efficient voltage pairs for a given $T_{amb}$.

Fig. \ref{fig:mkDelayWorker}(b) compares the total power consumption of the proposed technique and baseline.
Each curve shows the lower and upper bound of the power, where lower bound corresponds to $\alpha=0.1$ and upper bound corresponds to maximum dynamic power consumption, i.e., $\alpha=1.0$.
As explained above and showed in Fig. \ref{fig:activity}, power does not increase linearly with activity (i.e., upper bound of power is not 10$\times$ of lower bound) because leakage power is independent of activity and also activities of internal nodes are not linearly correlated with primary inputs activity.
As expected, lower temperatures have more power saving as there is more margin to reduce voltages.
Also, although the proposed method optimizes the voltages according to worst-case activity, our method still significantly improves power when activity is low.
Note that the baseline has fixed voltages; however, it also consumes less leakage power in lower temperature so its total power also reduces in lower temperatures.
It is noteworthy that in our experiments we observed the leakage power has an exponential relation of $e^{0.015T}$ with temperature, which is comparative to $e^{0.017T}$ we derived for Intel devices \cite{intelpower}.
The junction temperature of baseline exceeded 100\,$\degree$C when ambient temperature reaches $85\degree$C. Thus, Fig. \ref{fig:mkDelayWorker}(b) and (c) are limited to 85\,$\degree$C.

Finally, Fig. \ref{fig:mkDelayWorker}(c) shows the upper ($\alpha=1.0$) and lower ($\alpha=0.1$) bounds of increase in junction temperature of device tiles for different ambient temperatures (and corresponding voltages). Higher activity consumes more dynamic power hence has a higher impact on temperature.
There is a close correlation between Fig. \ref{fig:mkDelayWorker}(b) and (c) as steady-state junction temperature is correlated to total power, especially when design activity is uniform.
Please note that overlapping (almost) of lower bound of the proposed method with upper bound of the baseline is haphazard and specific to this benchmark.

\begin{table}[t]
\centering
\caption{Iterations of Algorithm \ref{algorithm1} on $\mathtt{mkDelayWorker}$ at $T_{amb}=\,$60\,$\degree$C.} \label{tab:algorithm1}
\large
\resizebox{0.46\textwidth}{!}{
\begin{tabular}{llllll}
\hline
Iter. & $V_{core} (mV)$ & $V_{bram} (mV)$ & Power ($mW$) & $T_{junct} (\degree C)$ & Time $(s)$ \\ \hline
$1$     & $740$        & $920$        & $485$        & $65.82$  & $10.9$     \\
$2$     & $750$        & $900$        & $558$        & $66.69$  & $3.1$      \\
$3$     & $750$        & $910$        & $564$        & $66.76$  & $3.1$      \\
$4$     & $750$        & $910$        & $564$        & $66.77$  & $3.1$      \\
$5$     & $750$        & $910$        & $564$        & $66.77$  & 3.1      \\ \hline
\end{tabular}
}
\normalsize
\end{table}
\normalsize

\textbf{$\bullet$ Algorithm Runtime.}
For all of our benchmarks, the flow converges in less than 6 iterations. At low $T_{amb}$ values, due to weak temperature-leakage feedback, the algorithm converges in 2--3 iterations.
On a typical desktop system, the first iteration takes less than 12 seconds, and in subsequent iterations the algorithm limits the search space to the boundary of the current solution, making each iteration less than 4 seconds.
Thermal simulation takes {\raise.17ex\hbox{$\scriptstyle\mathtt{\sim}$}}2.5 seconds of each iteration.
Table \ref{tab:algorithm1} shows the details of the static voltage scaling algorithm.
At first round, voltages are set to (0.74,\,0.92).
The resultant power increases the temperature by 5.82\,$\degree$C, which increases the delay (tightens the margin) and leakage.
Thus, the second iteration changes the voltages to (0.75,\,0.90) for timing closure.
The increase of temperature in the first iteration also considerably increases the (leakage) power, from 485\,mW to 558\,mW.
The temperature then starts converging; hence the subsequent voltage and power changes are insignificant.

\textbf{$\bullet$ Discussion.}
We do not change the voltages of other power rails such as auxiliary supply voltage ($V_{aux}$) and I/O voltage ($V_{io}$) as they enable interfacing with other devices and have relatively low power consumption.
We also do not touch the voltage of configuration SRAM cells as they use high-threshold mid-oxide transistors with two orders of magnitude less leakage \cite{tuan200690nm}.
In addition, we observed that reducing the voltage of SRAM cells causes voltage drop in pass-gate based multiplexer structure of resources, which increases the leakage power of buffers due to non-ideal voltage at their input.

\subsection{Proposed Thermal-Aware Energy Optimization Flow} \label{subsec:min-energy}

While performance is the major concern of high-end FPGA designs, total \textit{energy} usage is a primary concern of battery-backed and IoT applications.
The goal of optimal energy exploration is to find the voltage(s) $V_{opt}$ and clock period $d_{opt}$, for which $E(V_{opt}, d_{opt}) = P(V_{opt}, d_{opt}) \times d_{opt}$ is minimum, where $E(V_{opt}, d_{opt})$ is the design energy consumption rate operating with $V_{opt}$ and clock $d_{opt}$.
To obtain the $V_{opt}$ and clock period $d_{opt}$, clearly, the operating voltage $V_{opt}$ must be able to deliver the clock period of $d_{opt}$ to avoid timing violations.
Second, the design must operate with maximum possible frequency for the given voltage.
Otherwise, if the clock period is set to $\alpha \cdot d_{opt}$ ($\alpha > 1$), total energy becomes:

\small
\begin{align*}
E(V_{opt}, \boldsymbol{\alpha} \cdot d_{opt}) = \big(P_{lkg}(V_{opt}) + \frac{P_{dyn}(V_{opt})}{\boldsymbol{\alpha}}\big) \times \boldsymbol{\alpha} \cdot d_{opt} \quad \; \; (1) \\
= (\boldsymbol{\alpha} P_{lkg}(V_{opt}) + P_{dyn}(V_{opt})) d_{opt} > (P_{lkg}(V_{opt}) + P_{dyn}(V_{opt})) d_{opt}
\end{align*}
\normalsize
That is, scaling the clock by $\alpha$ scales the dynamic power by $\frac{1}{\alpha}$, but since the execution time also scales by $\alpha$, the total consumed dynamic energy remains the same.
Nonetheless, the leakage power is independent of the clock period. Thus, the leakage energy scales by $\alpha$.
Therefore, for a given voltage (which we aim to find the best one), the clock \textit{period} needs to be minimum possible to minimize total energy.

\begin{algorithm}[!t]
\small
\KwIn{$netlist$: Placed and routed design}
\KwIn{$T_{amb}$: Ambient temperature}
\KwIn{$\overrightarrow{\alpha}$: Input activities / sample inputs}

$E_{min} = \infty$\\
\For {$\forall\ V_{core},\ \forall\ V_{bram}$}{
$\overrightarrow{T}_{n \times n} = [T_{amb}, \cdots, T_{amb}]$\\
$\overrightarrow{\Delta T}_{n \times n} = [\infty, \cdots, \infty]$ \\
\While{$\| \overrightarrow{\Delta T} \|_{\infty} > \delta_T$}{
$d_{max} = \mathcal{T}(netlist, \overrightarrow{T}, V_{core}, V_{bram})$\\
$\overrightarrow{P}_{total} = \overrightarrow{P}_{lkg}(\overrightarrow{T}, V_{core}, V_{bram}) \ + $\\
$\qquad \qquad \ \ \overrightarrow{P}_{dyn}(netlist, \overrightarrow{\alpha}, d_{max}, V_{core}, V_{bram})$\\
$\overrightarrow{T}_{old} = \overrightarrow{T}$\\
$\overrightarrow{T} = HotSpot(\overrightarrow{P}_{lkg} + \overrightarrow{P}_{dyn})$ \\
$\overrightarrow{\Delta T} = \overrightarrow{T} - \overrightarrow{T}_{old}$\\
}
\If {$d_{max} \times \sum_{i}{\overrightarrow{P_i}} < E_{min}$}{
$E_{min} = d_{max} \times \sum_{i}{\overrightarrow{P_i}}$\\
$V_{core_{min}} = V_{core}$\\
$V_{bram_{min}} = V_{bram}$\\
}
}
\Return $V_{core_{min}}, V_{bram_{min}}$
\caption{{\sc} Thermal-Aware Energy Optimization} \label{algorithm2}
\end{algorithm}
\normalsize

That being said, we derive the new Algorithm \ref{algorithm2} that looks for the ($V_{core}$, $V_{bram}$) pair that achieves minimum power-delay product as the energy metric whilst also exploits the temperature headroom for further efficiency.
Clearly, having the temperature-delay-voltage and voltage-power characterization is vital for the reliability and efficiency of the proposed flow.
Having Algorithm \ref{algorithm1} already explained, understanding the Algorithm \ref{algorithm2} is straightforward.
Essentially, it looks for all $(V_{core}, V_{bram})$ pairs, and as reasoned above, finds the maximum frequency considering thermal margin.
In contrast with the voltage scaling flow, Algorithm \ref{algorithm2} exploits the available thermal margin to maximize the frequency for a candidate voltage, rather than lowering the voltage for a fixed frequency.
Thermal simulation (line 10) is again crucial as changing the frequency (line 6) changes the power and hence temperature.

Algorithm \ref{algorithm1} was performing thermal simulation for the best $(V_{core}, V_{bram})$ at each iteration and we observed that, in the worst case, it converges in less than eight iterations.
However, Algorithm \ref{algorithm2} needs to explore all $|V_{core}| \times |V_{bram}|$ combinations and perform several thermal simulations under each.
It could take up to several hours for large benchmarks.
We enhanced it by, first, skipping a $(V_{core}, V_{bram})$ combination if its energy in the initial loop (i.e., before involving temperature-delay feedback of line 10) was larger than the already found optimum.
In addition, we also considered a small temperature margin of 0.1\,$\degree$C, so we could avoid the thermal simulation of cases with power within $\frac{0.1}{\theta_{JA}}$ range of a previously obtained case.
These optimizations reduced the average runtime on benchmarks by two-order of magnitude (from 72 minutes to 49 seconds) with virtually no impact on the solution.

\subsection{Timing-Speculative Voltage Over-Scaling} \label{subsec:overscaling}
Timing speculation has been shown to provide opportunistic power reduction in applications that can inherently tolerate a certain amount of error. Examples are: (a) image processing circuits such as DCT/IDCT where small drop of PSNR (Peak Signal to Noise Ratio) might not be perceived by human \cite{amrouch2016reliability}. (b) Deep Neural Networks (DNNs) because of their pooling layers that filter out a significant portion of intermediate results and also the stochastic nature of the gradient descent \cite{zhang2018thundervolt} (c) light-weight alternatives of DNNs such as the brain-inspired computing that performs the main machine learning applications by using inexpensive operations on hypervectors that can bear a certain amount of inaccuracy \cite{imani2017exploring}, etc.
Timing-speculative voltage over-scaling is orthogonal to the thermal-aware voltage reduction with the opportunity of violating the timing for more significant power saving.

Timing-speculative voltage scaling requires post P\&R simulation of the targeted design with the timing data at the scaled voltage to estimate the incurred inaccuracy.
In standard-cell design approaches, timing speculation can be realized by generating the same cell library under scaled voltages or providing multiple operating condition modes at the same library.
Thus, the synthesized design can be simulated using the new timing data.
Nonetheless, to the best our knowledge, there is no FPGA framework for post place and route (timing) simulation, particularly under varying voltages.

\begin{figure}[t]
  \centering
  \includegraphics[width=0.49\textwidth, height=2.4cm]{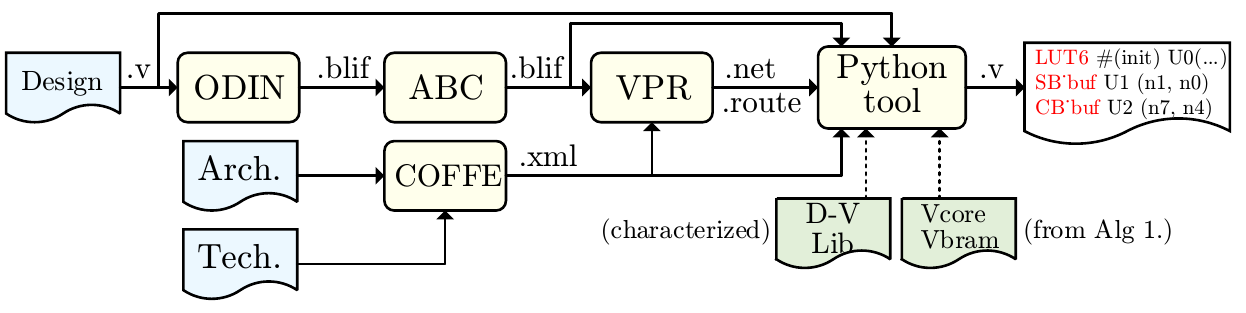}
  \caption{The proposed simulation flow for FPGA-mapped applications, enabling speculative voltage scaling.}
  \label{fig:flow}
\end{figure}

By taking advantage of our characterization libraries, we build our simulation framework upon the Verilog-to-Routing (VTR) \cite{luu2014vtr} project explained in Section \ref{subsec:preliminary}.
Fig. \ref{fig:flow} demonstrates the proposed post P\&R simulation framework.
The flow begins with synthesis (using ODIN \cite{jamieson2010odin}) and technology mapping (using Berkeley ABC \cite{brayton2010abc}) of the HDL description to BLIF format that can be imported to the VPR P\&R tool \cite{luu2014vtr}.
Architectural description of the target FPGA needed by VPR can be hand-written or auto-generated by COFFE \cite{chiasson2013coffe}.
VPR generates a $\mathtt{.net}$ file that describes the placement information of resources, and a $\mathtt{.route}$ file that provides the routing information of design nets.
Finally, we developed a Python-based tool to analyze VPR outputs and instantiate the FPGA's utilized components using primitives similar to Xilinx syntax in Verilog HDL.
The generated file is augmented with the delay of the resources obtained by parsing the VPR outputs.
It is noteworthy that all configurable multiplexers (SBs, CBs, etc.) are programmed in a place and routed design, so we could simplify their functionality as a buffer just to incorporate their delay information.
VPR's placement output ($\mathtt{.net}$) does not include a functional description. Thus, we retrieve the configuration of LUTs from the technology-mapped BLIF.
Similarly, BLIF files do not contain BRAMs initialization, so we read them back from the original HDL file.
To make the voltage over-scaling efficient, we utilize the proposed thermal-aware voltage scaling as follows.
For a given timing rate (e.g., 1.1$\times$ of original clock), we change the timing condition of Algorithm \ref{algorithm1} (line 7) to meet the new constraint (1.1$\times$ of $d_{worst}$). Hence, the obtained over-scaled voltages are optimal for that allowed amount of violation.
We repeat it for different timing violation rates.
In Section \ref{sec:res} we report the additional power saving granted by speculative voltage-scaling.

\section{Experimental Results} \label{sec:res}

\begin{figure}[t]
\centering
\subfloat[Range of power reduction at 40\,$\degree$C (left axis) and corresponding core and memory voltages of each benchmark (right axis). \label{fig:power_saving_40}]
        {\includegraphics[width=0.45\textwidth]{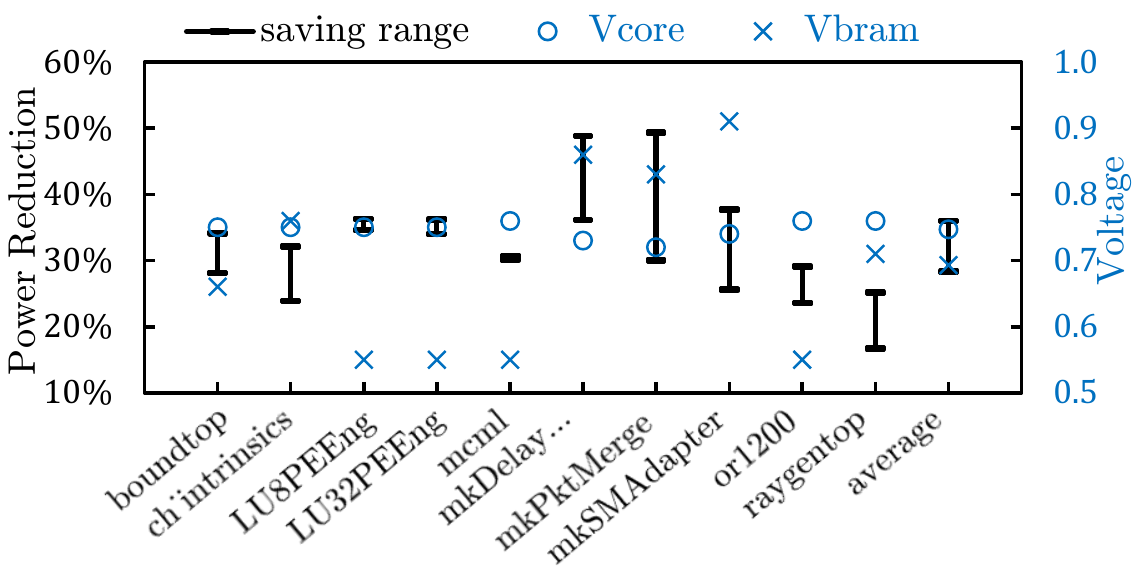}}

\subfloat[Range of power reduction at 65\,$\degree$C (left axis) and corresponding core and memory voltages of each benchmark (right axis). \label{fig:power_saving_65}]
         {\includegraphics[width=0.45\textwidth]{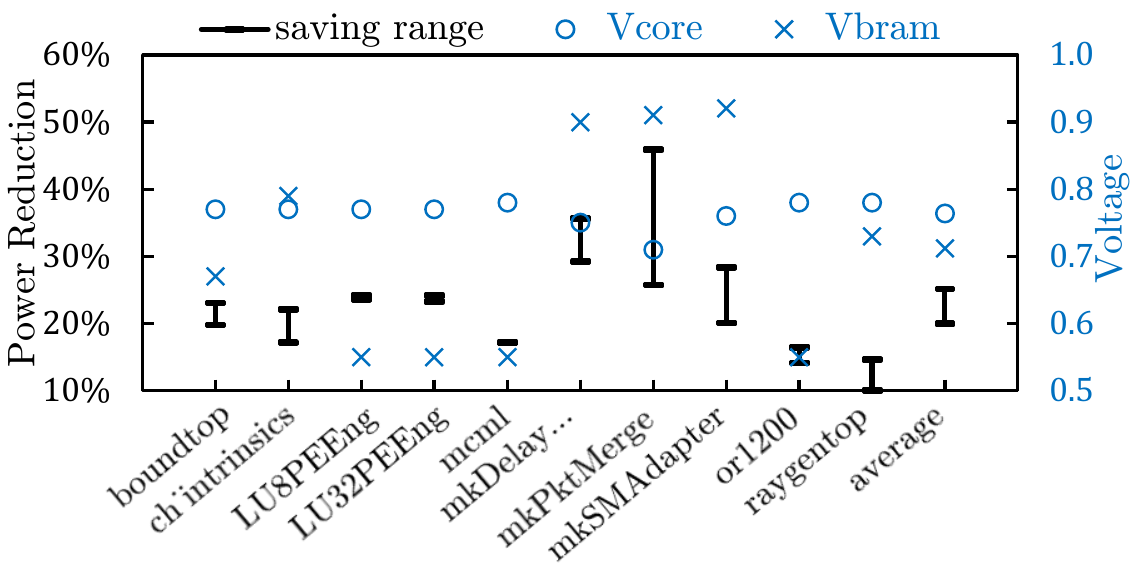}}         

\caption[Motivational Figure]
        {Power reduction and voltages for 40\,$\degree$C and 65\,$\degree$C board temperatures.}
    \label{fig:power_saving}
\end{figure}

\textbf{$\bullet$ Power Reduction.}
Fig. \ref{fig:power_saving} demonstrates the power reduction using the proposed flow of Algorithm \ref{algorithm1}.
The flow finds the optimum core and memory voltage pair ($V_{core}$ and $V_{bram}$) assuming highest inputs activity ($\alpha$) to guarantee it satisfies temperature corners.
In practice, however, the input activity range might be lower.
Therefore, for the obtained optimal voltages, we assumed a varying activity $\alpha \in$ [0.1, 1.0] and calculated the power reduction for the entire range.
Therefore, Fig. \ref{fig:power_saving} demonstrates a range of power saving.
The right axis of this figure also shows the optimal $V_{core}$ and $V_{bram}$ voltages for each benchmark.
In several benchmarks, the paths containing memories were significantly shorter than critical paths.
For instance, in $\mathtt{LU8PEEng}$, the critical path is 21$\times$ longer than the longest BRAM path.
For these paths, $V_{bram}$ is reduced down to 0.55\,V, which we set as the lowest voltage level before device crashes \cite{salami2018bram}.
In Fig. \ref{fig:power_saving}(a) we considered a device operating at $T_{amb} = $\,40\,$\degree$C with
$\theta_{JA} = 12\,\sfrac{\degree C}{W}$, and in Fig. \ref{fig:power_saving}(b) as considered more high-end device operating at 65\,$\degree$C with $\theta_{JA} = 2\,\sfrac{\degree C}{W}$ (see Section \ref{subsec:preliminary} for details of experiments).
The opportunity of power saving reduces in higher temperatures.
At 40\,$\degree$C, the average power saving of 10 benchmarks is 28.3\%--36.0\% (depends on activity), while at 65\,$\degree$C it becomes 20.0\%--25.0\%.
We observed up to 9.2\,$\degree$C increase in the junction temperature of the baseline, which reduced to 5.9\,$\degree$C in the proposed method due to consuming less power.
Benchmarks have different power reduction and optimal voltages based on the resources on critical paths (that determine the voltage scaling limit), used resources, the activity of nodes, etc.
Comparing Fig. \ref{fig:power_saving}(a) and (b) also reveals how differently the voltages of benchmarks need to be adjusted moving from 40\,$\degree$C to 65\,$\degree$C: $\mathtt{raygentop}$ needs boosting both voltages by 20\,mV, $\mathtt{or1200}$ needs only $+$20\,mV of core rail, and $\mathtt{mkPktMerge}$ needs 80\,mV increase of memory rail with 10\,mV reduction of core voltage, suggesting the necessity of dynamic implementation (see Section \ref{subsec:flow}) for best efficacy.

\begin{figure}[t]
  \centering
  \includegraphics[width=0.45\textwidth]{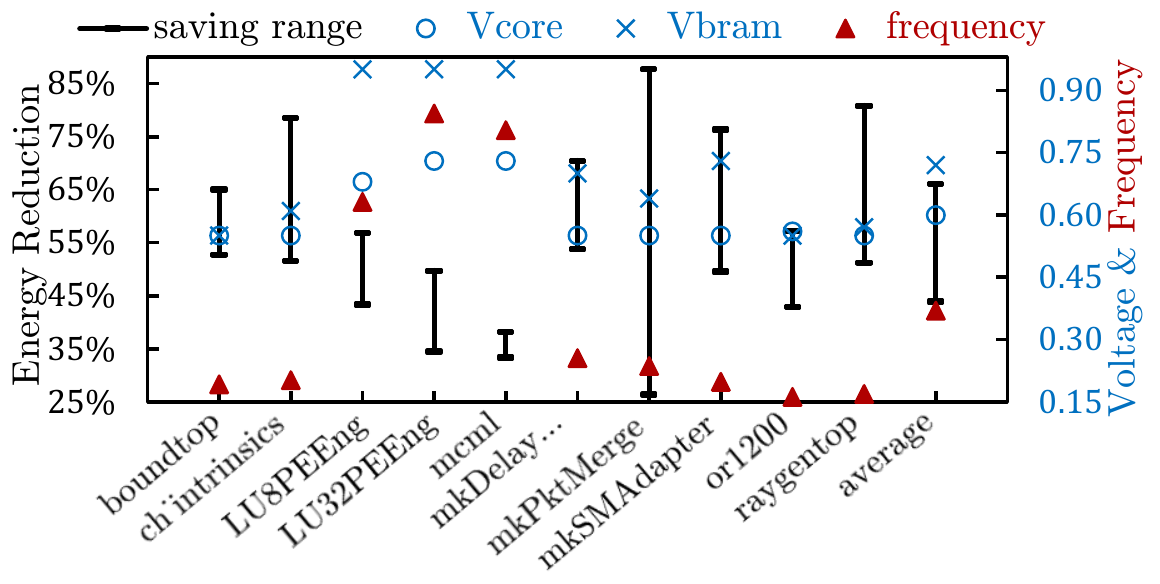}
  \caption{Range of energy savings at 65\,$\degree$C (left axis), and corresponding optimal voltage values and frequency ratio (right axis).}
  \label{fig:energy_saving}
\end{figure}

\textbf{$\bullet$ Energy Reduction.}
Fig. \ref{fig:energy_saving} shows the range of energy reduction (left axis) of each benchmark using the proposed energy optimization flow at 65\,$\degree$C.
The points (\textcolor{blue}{$\boldsymbol{\times}$}, \textcolor{blue}{$\boldsymbol{\circ}$}, \textcolor{Red}{$\boldsymbol{\blacktriangle}$}) correspond to the right axis and show the optimal voltage values and frequency ratio.
Remember that the goal of our energy minimization flow was to find out the minimum energy consumption point (power-delay product) by compromising the delay and power, so the delay has been increased by $\frac{1}{0.37}=2.7\times$ while the overall consumed energy is improved by 44\%--66\% depending on input activities.
There are obvious differences with optimal points of power and energy reduction flows.
Unlike the power flow, here, $V_{bram}$ of above-mentioned benchmarks (e.g., $\mathtt{LU8PEEng}$) have not been shrunk down to 0.55\,V because the delay of their critical paths is also increased, hence the memory voltage cannot be freely reduced.
The ranges of energy savings are also more stretched (e.g., in $\mathtt{mkPktMerge}$) because in higher activities, memory dynamic energy becomes the dominant contributor to total energy consumption, hence, throttling its energy becomes worthwhile even considering the increased delay.
As it can be seen in the figure, its $V_{core}$ is reduced to 0.64\,V while in power reduction flow (Fig. \ref{fig:power_saving}(b)) it could be reduced to 0.91\,V because the clock delay must have been remained fixed.

\begin{figure}[t]
  \centering
  \includegraphics[width=0.48\textwidth]{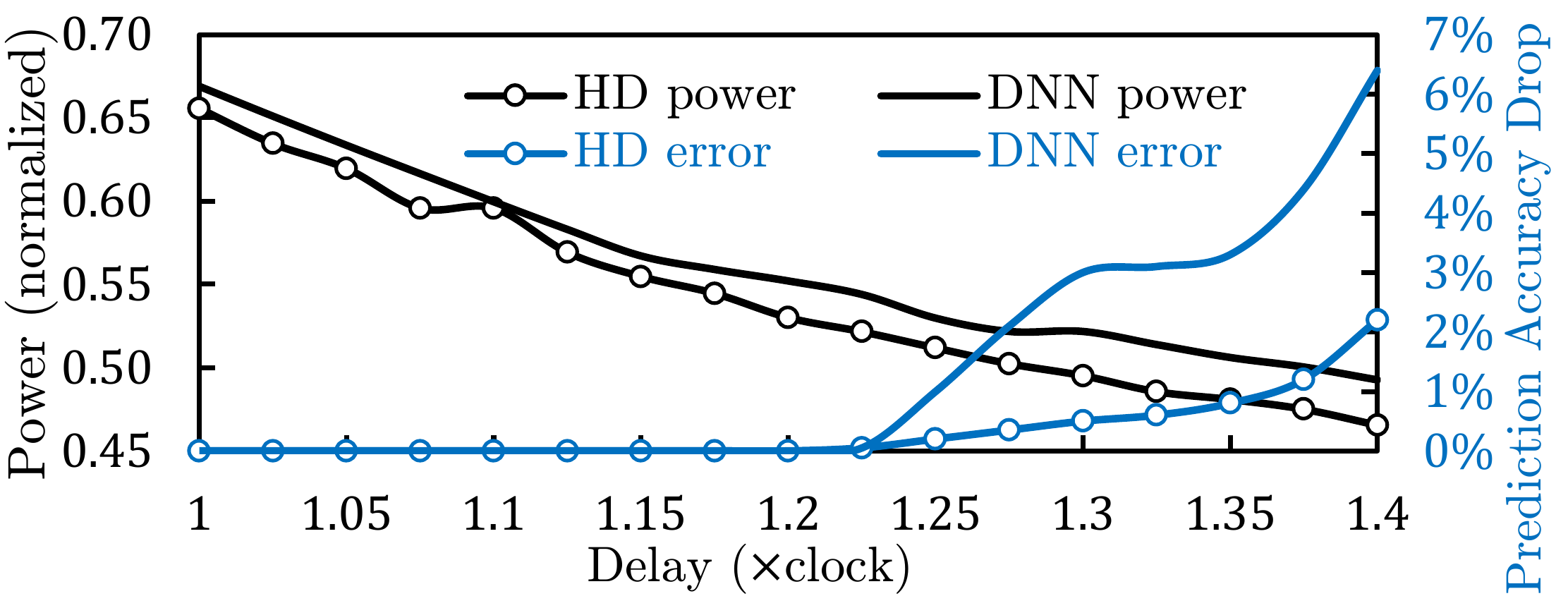}
  \caption{Power reduction (left axis) and error increase (right axis) under voltage over-scaling. X-axis shows violation of critical path delay. $T_{amb}$ is 40\,$\degree$C.}
  \label{fig:over-scaling}
\end{figure}

\textbf{$\bullet$ Speculative Voltage Over-Scaling.}
We chose $\mathtt{LeNet}$ \cite{lecun1998gradient} as a classic CNN for handwritten digit recognition and implemented as a systolic array architecture \cite{zhang2018fate}.
Its relatively small size makes the timing simulation computationally tractable.
We also selected another machine learning algorithm based on computing with hyperdimensional ($\mathtt{HD}$) \cite{schmuck2018hardware} vectors to detect two face/non-face classes among 10,000 web faces of a face detection dataset ($\mathtt{FACE}$) from Caltech \cite{griffin2007caltech}.
Fig. \ref{fig:over-scaling} shows the result of thermal-aware voltage over-scaling.
Initially, CP delay is $1\times$ of original clock, meaning that no timing violation is allowed and the {\raise.17ex\hbox{$\scriptstyle\mathtt{\sim}$}}34\% power reduction is because of thermal-aware voltage scaling.
Thereafter, we allow up to 40\% violation of CP delay, where accuracy drop becomes noticeable at 1.2$\times$ of the original clock.
This tolerance is because DNNs are intrinsically error-tolerant (e.g., allow quantization of weights down to three bits in the $\mathtt{LeNet}$ \cite{elthakeb2018releq}).
Similarly, previous studies of $\mathtt{HD}$ have shown an accuracy drop of merely 4\% when up to 30\% of the vectors bits are flipped (i.e., noisy) \cite{imani2017exploring} mainly because orthogonality of vectors, making them discernible under error.
Based on Fig. \ref{fig:over-scaling}, when voltage over-scaling increases the CP delay to 1.35$\times$ of the clock period, errors start spiking.
At this point, by respectively 3\% and 0.5\% accuracy drop, $\mathtt{LeNet}$ and $\mathtt{HD}$ powers are reduced by 48\% and 50\%, which means additional 15\% and 16\% improvement compared to our original voltage-scaling (with {\raise.17ex\hbox{$\scriptstyle\mathtt{\sim}$}}34\% improvement for both) in which CP delay does not exceed clock period.

\section{Conclusion} \label{sec:conc}
In this paper, we proposed power and energy optimization techniques for FPGAs by characterizing FPGA resources and utilizing thermal headroom.
Keeping the performance intact, the voltage scaling flow determines the optimal voltages of designs based on their thermal distribution and can be implemented statically with fixed voltage(s), or dynamically using programmable voltage regulators at highly varying ambient temperatures.
Our proposed energy optimization flow compromises the delay and power to seek optimal point that minimizes total consumed energy.
Finally, we proposed a timing simulation framework that provides further power saving opportunity by making the impact of voltage over-scaling observable.

\section*{Acknowledgements}
This work was partially supported by CRISP, one of six centers in JUMP, an SRC program sponsored by DARPA, and also NSF grants \#1911095, \#1826967, \#1730158 and \#1527034.

\bibliographystyle{IEEEtran}
\bibliography{references} 

\begin{thebibliography}{10}
\providecommand{\url}[1]{#1}
\csname url@samestyle\endcsname
\providecommand{\newblock}{\relax}
\providecommand{\bibinfo}[2]{#2}
\providecommand{\BIBentrySTDinterwordspacing}{\spaceskip=0pt\relax}
\providecommand{\BIBentryALTinterwordstretchfactor}{4}
\providecommand{\BIBentryALTinterwordspacing}{\spaceskip=\fontdimen2\font plus
\BIBentryALTinterwordstretchfactor\fontdimen3\font minus
  \fontdimen4\font\relax}
\providecommand{\BIBforeignlanguage}[2]{{%
\expandafter\ifx\csname l@#1\endcsname\relax
\typeout{** WARNING: IEEEtran.bst: No hyphenation pattern has been}%
\typeout{** loaded for the language `#1'. Using the pattern for}%
\typeout{** the default language instead.}%
\else
\language=\csname l@#1\endcsname
\fi
#2}}
\providecommand{\BIBdecl}{\relax}
\BIBdecl

\bibitem{putnam2014reconfigurable}
A.~Putnam, A.~M. Caulfield, E.~S. Chung, D.~Chiou, K.~Constantinides, J.~Demme
  \emph{et~al.}, ``A reconfigurable fabric for accelerating large-scale
  datacenter services,'' \emph{ACM SIGARCH Computer Architecture News},
  vol.~42, no.~3, pp. 13--24, 2014.

\bibitem{lacey2016deep}
G.~Lacey, G.~W. Taylor, and S.~Areibi, ``Deep learning on fpgas: Past, present,
  and future,'' \emph{arXiv preprint arXiv:1602.04283}, 2016.

\bibitem{stratix10}
``Intel stratix 10 logic array blocks and adaptive logic modules user guide,''
  User Guide, Intel, September 2017.

\bibitem{ahmed2018automatic}
I.~Ahmed, S.~Zhao, J.~Meijers, O.~Trescases, and V.~Betz, ``Automatic bram
  testing for robust dynamic voltage scaling for fpgas,'' in \emph{2018 28th
  International Conference on Field Programmable Logic and Applications
  (FPL)}.\hskip 1em plus 0.5em minus 0.4em\relax IEEE, 2018, pp. 68--687.

\bibitem{shen2019fast}
L.~L. Shen, I.~Ahmed, and V.~Betz, ``Fast voltage transients on fpgas: Impact
  and mitigation strategies,'' in \emph{2019 IEEE 27th Annual International
  Symposium on Field-Programmable Custom Computing Machines (FCCM)}.\hskip 1em
  plus 0.5em minus 0.4em\relax IEEE, 2019, pp. 271--279.

\bibitem{blaauw2014iot}
D.~Blaauw, D.~Sylvester, P.~Dutta, Y.~Lee, I.~Lee, S.~Bang \emph{et~al.}, ``Iot
  design space challenges: Circuits and systems,'' in \emph{Symposium on VLSI
  Technology (VLSI-Technology): Digest of Technical Papers}.\hskip 1em plus
  0.5em minus 0.4em\relax IEEE, 2014, pp. 1--2.

\bibitem{venkataramani2016efficient}
S.~Venkataramani, K.~Roy, and A.~Raghunathan, ``Efficient embedded learning for
  iot devices,'' in \emph{Design Automation Conference (ASP-DAC), 21st Asia and
  South Pacific}.\hskip 1em plus 0.5em minus 0.4em\relax IEEE, 2016, pp.
  308--311.

\bibitem{xilinx7}
``Lowering power at 28 nm with xilinx 7 series devices,'' White Paper, Xilinx,
  January 2015.

\bibitem{bao2009line}
M.~Bao, A.~Andrei, P.~Eles, and Z.~Peng, ``On-line thermal aware dynamic
  voltage scaling for energy optimization with frequency/temperature dependency
  consideration,'' in \emph{Design Automation Conference, 46th ACM/IEEE}.\hskip
  1em plus 0.5em minus 0.4em\relax IEEE, 2009, pp. 490--495.

\bibitem{lefurgy2011active}
C.~R. Lefurgy, A.~J. Drake, M.~S. Floyd, M.~S. Allen-Ware, B.~Brock, J.~A.
  Tierno \emph{et~al.}, ``Active management of timing guardband to save energy
  in power7,'' in \emph{Microarchitecture (MICRO), 44th Annual IEEE/ACM
  International Symposium on}.\hskip 1em plus 0.5em minus 0.4em\relax IEEE,
  2011, pp. 1--11.

\bibitem{amrouch2018voltage}
H.~Amrouch, B.~Khaleghi, and J.~Henkel, ``Voltage adaptation under temperature
  variation,'' in \emph{15th International Conference on Synthesis, Modeling,
  Analysis and Simulation Methods and Applications to Circuit Design
  (SMACD)}.\hskip 1em plus 0.5em minus 0.4em\relax IEEE, 2018, pp. 57--60.

\bibitem{lewis2005stratix}
D.~Lewis, E.~Ahmed, G.~Baeckler, V.~Betz, M.~Bourgeault, D.~Cashman
  \emph{et~al.}, ``The stratix ii logic and routing architecture,'' in
  \emph{Proceedings of the ACM/SIGDA 13th international symposium on
  Field-programmable gate arrays}.\hskip 1em plus 0.5em minus 0.4em\relax ACM,
  2005, pp. 14--20.

\bibitem{seifoori2018introduction}
Z.~Seifoori, Z.~Ebrahimi, B.~Khaleghi, and H.~Asadi, ``Introduction to emerging
  sram-based fpga architectures in dark silicon era,'' \emph{Advances in
  Computers}, 2018.

\bibitem{levine2014dynamic}
J.~M. Levine, E.~Stott, and P.~Y. Cheung, ``Dynamic voltage \& frequency
  scaling with online slack measurement,'' in \emph{Proceedings of the 2014
  ACM/SIGDA international symposium on Field-programmable gate arrays}.\hskip
  1em plus 0.5em minus 0.4em\relax ACM, 2014, pp. 65--74.

\bibitem{nunez2017adaptive}
J.~Nunez-Yanez, ``Adaptive voltage scaling in a heterogeneous fpga device with
  memory and logic in-situ detectors,'' \emph{Microprocessors and
  Microsystems}, vol.~51, pp. 227--238, 2017.

\bibitem{zhao2018robust}
S.~Zhao, I.~Ahmed, C.~Lamoureux, A.~Lotfi, V.~Betz, and O.~Trescases, ``Robust
  self-calibrated dynamic voltage scaling in fpgas with thermal and ir-drop
  compensation,'' \emph{IEEE Transactions on Power Electronics}, vol.~33,
  no.~10, pp. 8500--8511, 2018.

\bibitem{amouri2013accurate}
A.~Amouri, H.~Amrouch, T.~Ebi, J.~Henkel, and M.~Tahoori, ``Accurate
  thermal-profile estimation and validation for fpga-mapped circuits,'' in
  \emph{Field-Programmable Custom Computing Machines (FCCM), 2013 IEEE 21st
  Annual International Symposium on}.\hskip 1em plus 0.5em minus 0.4em\relax
  IEEE, 2013, pp. 57--60.

\bibitem{chiasson2013should}
C.~Chiasson and V.~Betz, ``Should fpgas abandon the pass-gate?'' in \emph{FPL},
  vol.~13, 2013, pp. 1--8.

\bibitem{salami2018bram}
B.~Salami, O.~S.~Unsal, and A.~Cristal~Kestelman, ``Comprehensive evaluation of
  supply voltage underscaling in fpga on-chip memories,'' in \emph{Proceedings
  of the 51th Annual IEEE/ACM International Symposium on
  Microarchitecture}.\hskip 1em plus 0.5em minus 0.4em\relax ACM, 2018.

\bibitem{khaleghi2019estimating}
B.~Khaleghi, B.~Omidi, H.~Amrouch, J.~Henkel, and H.~Asadi, ``Estimating and
  mitigating aging effects in routing network of fpgas,'' \emph{IEEE
  Transactions on Very Large Scale Integration (VLSI) Systems}, vol.~27, no.~3,
  pp. 651--664, 2019.

\bibitem{khaleghi2019thermal}
B.~Khaleghi and T.~{\v{S}}. Rosing, ``Thermal-aware design and flow for fpga
  performance improvement,'' in \emph{2019 Design, Automation \& Test in Europe
  Conference \& Exhibition (DATE)}.\hskip 1em plus 0.5em minus 0.4em\relax
  IEEE, 2019, pp. 342--347.

\bibitem{salamat2019workload}
S.~Salamat, B.~Khaleghi, M.~Imani, and T.~Rosing, ``Workload-aware
  opportunistic energy efficiency in multi-fpga platforms,'' \emph{arXiv
  preprint arXiv:1908.06519}, 2019.

\bibitem{chiasson2013coffe}
C.~Chiasson and V.~Betz, ``Coffe: Fully-automated transistor sizing for
  fpgas,'' in \emph{Field-Programmable Technology (FPT), 2013 International
  Conference on}.\hskip 1em plus 0.5em minus 0.4em\relax IEEE, 2013, pp.
  34--41.

\bibitem{ptm}
\BIBentryALTinterwordspacing
Predictive technology model. [Online]. Available: \url{http://ptm.asu.edu/}
\BIBentrySTDinterwordspacing

\bibitem{yazdanshenas2017don}
S.~Yazdanshenas, K.~Tatsumura, and V.~Betz, ``Don't forget the memory:
  Automatic block ram modelling, optimization, and architecture exploration,''
  in \emph{Proceedings of the 2017 ACM/SIGDA International Symposium on
  Field-Programmable Gate Arrays}.\hskip 1em plus 0.5em minus 0.4em\relax ACM,
  2017, pp. 115--124.

\bibitem{tuan200690nm}
T.~Tuan, S.~Kao, A.~Rahman, S.~Das, and S.~Trimberger, ``A 90nm low-power fpga
  for battery-powered applications,'' in \emph{Proceedings of the 2006
  ACM/SIGDA 14th international symposium on Field programmable gate
  arrays}.\hskip 1em plus 0.5em minus 0.4em\relax ACM, 2006, pp. 3--11.

\bibitem{stratixiv}
``Stratix iv device handbook,'' Datasheet, Intel, September 2014.

\bibitem{nangate2008california}
\BIBentryALTinterwordspacing
Nangate open cell library. [Online]. Available: \url{http://nangate.com/}
\BIBentrySTDinterwordspacing

\bibitem{xpower}
``Xilinx power estimator user guide,'' User Guide, Xilinx, 2018.

\bibitem{luu2014vtr}
J.~Luu, J.~Goeders, M.~Wainberg, A.~Somerville, T.~Yu, K.~Nasartschuk
  \emph{et~al.}, ``Vtr 7.0: Next generation architecture and cad system for
  fpgas,'' \emph{ACM Transactions on Reconfigurable Technology and Systems
  (TRETS)}, vol.~7, no.~2, p.~6, 2014.

\bibitem{lewis2009architectural}
D.~Lewis, E.~Ahmed, D.~Cashman, T.~Vanderhoek, C.~Lane, A.~Lee \emph{et~al.},
  ``Architectural enhancements in stratix-iii™ and stratix-iv™,'' in
  \emph{Proceedings of the ACM/SIGDA international symposium on Field
  programmable gate arrays}.\hskip 1em plus 0.5em minus 0.4em\relax ACM, 2009,
  pp. 33--42.

\bibitem{lamoureux2006activity}
J.~Lamoureux and S.~J. Wilton, ``Activity estimation for field-programmable
  gate arrays,'' in \emph{Field Programmable Logic and Applications, 2006.
  FPL'06. International Conference on}.\hskip 1em plus 0.5em minus 0.4em\relax
  IEEE, 2006, pp. 1--8.

\bibitem{zhang2015hotspot}
R.~Zhang, M.~R. Stan, and K.~Skadron, ``Hotspot 6.0: Validation, acceleration
  and extension,'' \emph{University of Virginia, Tech. Rep}, 2015.

\bibitem{velusamy2005monitoring}
S.~Velusamy, W.~Huang, J.~Lach, M.~Stan, and K.~Skadron, ``Monitoring
  temperature in fpga based socs,'' in \emph{2005 International Conference on
  Computer Design}.\hskip 1em plus 0.5em minus 0.4em\relax IEEE, 2005, pp.
  634--637.

\bibitem{intelpower}
``Powerplay early power estimator user guide,'' User Guide, Intel, February
  2017.

\bibitem{timing}
``Timing closure user guide,'' User Guide, Xilinx, January 2012.

\bibitem{amrouch2017optimizing}
H.~Amrouch, B.~Khaleghi, and J.~Henkel, ``Optimizing temperature guardbands,''
  in \emph{Design, Automation \& Test in Europe Conference \& Exhibition
  (DATE)}.\hskip 1em plus 0.5em minus 0.4em\relax IEEE, 2017, pp. 175--180.

\bibitem{intelsensor}
``Intel fpga temperature sensor ip core user guide,'' User Guide, Intel, May
  2018.

\bibitem{burton2014fivr}
E.~A. Burton, G.~Schrom, F.~Paillet, J.~Douglas, W.~J. Lambert,
  K.~Radhakrishnan \emph{et~al.}, ``Fivr—fully integrated voltage regulators
  on 4th generation intel{\textregistered} core™ socs,'' in \emph{2014 IEEE
  Applied Power Electronics Conference and Exposition-APEC 2014}.\hskip 1em
  plus 0.5em minus 0.4em\relax IEEE, 2014, pp. 432--439.

\bibitem{tian2019temporal}
S.~Tian and J.~Szefer, ``Temporal thermal covert channels in cloud fpgas,'' in
  \emph{Proceedings of the 2019 ACM/SIGDA International Symposium on
  Field-Programmable Gate Arrays}.\hskip 1em plus 0.5em minus 0.4em\relax ACM,
  2019, pp. 298--303.

\bibitem{ebi2011economic}
T.~Ebi, D.~Kramer, W.~Karl, and J.~Henkel, ``Economic learning for
  thermal-aware power budgeting in many-core architectures,'' in
  \emph{Proceedings of the seventh IEEE/ACM/IFIP international conference on
  Hardware/software codesign and system synthesis}.\hskip 1em plus 0.5em minus
  0.4em\relax ACM, 2011, pp. 189--196.

\bibitem{amrouch2016reliability}
H.~Amrouch, B.~Khaleghi, A.~Gerstlauer, and J.~Henkel, ``Reliability-aware
  design to suppress aging,'' in \emph{2016 53nd ACM/EDAC/IEEE Design
  Automation Conference (DAC)}.\hskip 1em plus 0.5em minus 0.4em\relax IEEE,
  2016, pp. 1--6.

\bibitem{zhang2018thundervolt}
J.~Zhang, K.~Rangineni, Z.~Ghodsi, and S.~Garg, ``Thundervolt: enabling
  aggressive voltage underscaling and timing error resilience for energy
  efficient deep learning accelerators,'' in \emph{Proceedings of the 55th
  Annual Design Automation Conference}.\hskip 1em plus 0.5em minus 0.4em\relax
  ACM, 2018, p.~19.

\bibitem{imani2017exploring}
M.~Imani, A.~Rahimi, D.~Kong, T.~Rosing, and J.~M. Rabaey, ``Exploring
  hyperdimensional associative memory,'' in \emph{2017 IEEE International
  Symposium on High-Performance Computer Architecture (HPCA)}.\hskip 1em plus
  0.5em minus 0.4em\relax IEEE, 2017, pp. 445--456.

\bibitem{jamieson2010odin}
P.~A. Jamieson and K.~B. Kent, ``Odin ii: an open-source verilog hdl synthesis
  tool for fpga cad flows,'' in \emph{Proceedings of the 18th annual ACM/SIGDA
  international symposium on Field programmable gate arrays}.\hskip 1em plus
  0.5em minus 0.4em\relax ACM, 2010, pp. 288--288.

\bibitem{brayton2010abc}
R.~Brayton and A.~Mishchenko, ``Abc: An academic industrial-strength
  verification tool,'' in \emph{International Conference on Computer Aided
  Verification}.\hskip 1em plus 0.5em minus 0.4em\relax Springer, 2010, pp.
  24--40.

\bibitem{lecun1998gradient}
Y.~LeCun, L.~Bottou, Y.~Bengio, P.~Haffner \emph{et~al.}, ``Gradient-based
  learning applied to document recognition,'' \emph{Proceedings of the IEEE},
  vol.~86, no.~11, pp. 2278--2324, 1998.

\bibitem{zhang2018fate}
J.~J. Zhang and S.~Garg, ``Fate: fast and accurate timing error prediction
  framework for low power dnn accelerator design,'' in \emph{Proceedings of the
  International Conference on Computer-Aided Design}.\hskip 1em plus 0.5em
  minus 0.4em\relax ACM, 2018, p.~24.

\bibitem{schmuck2018hardware}
M.~Schmuck, L.~Benini, and A.~Rahimi, ``Hardware optimizations of dense binary
  hyperdimensional computing: Rematerialization of hypervectors, binarized
  bundling, and combinational associative memory,'' \emph{arXiv preprint
  arXiv:1807.08583}, 2018.

\bibitem{griffin2007caltech}
G.~Griffin, A.~Holub, and P.~Perona, ``Caltech-256 object category dataset,''
  2007.

\bibitem{elthakeb2018releq}
A.~T. Elthakeb, P.~Pilligundla, A.~Yazdanbakhsh, F.~Mireshghallah, and
  H.~Esmaeilzadeh, ``Releq: A reinforcement learning approach for deep
  quantization of neural networks,'' \emph{arXiv preprint arXiv:1811.01704},
  2018.

\end{thebibliography}

\end{document}